\documentclass[twocolumn,amsmath,showpacs,amssymb]{revtex4-1}
\usepackage{graphicx,bm}

\def\unitangstrom{\,\textrm{\AA}}
\def\unitev{~{\rm eV}}
\def\unitmev{\,{\rm meV}}
\def\unitry{\,{\rm Ry}}

\def\unitkelvin{\,{\rm K}}

\def\kprime{K^{\prime}}

\begin{document}

\title{Phonon-assisted indirect transitions in angle-resolved\\
  photoemission spectra of graphite and graphene}

\author{Pourya~Ayria$^1$}
\email[Electronic address: ]{pourya@flex.phys.tohoku.ac.jp}

\author{Shin-ichiro Tanaka$^2$}
\author{Ahmad R. T. Nugraha$^1$}
\author{Mildred S. Dresselhaus$^{3,4}$}
\author{Riichiro Saito$^1$}

\affiliation{$^1$Department of Physics, Tohoku University, Sendai
  980-8578, Japan\\
  $^2$The Institute of Scientific and Industrial Research, Osaka
  University, Osaka 567-0047, Japan\\
  $^3$Department of Electrical Engineering, Massachusetts Institute of
  Technology, Cambridge, MA 02139-4307, USA\\ $^4$Department of
  Physics, Massachusetts Institute of Technology, Cambridge, MA
  02139-4307, USA}

\date{\today}

\begin{abstract}
  Indirect transitions of electrons in graphene and graphite are
  investigated by means of angle-resolved photoemission
  spectroscopy~(ARPES) with several different incident photon energies
  and light polarizations.  The theoretical calculations of the
  indirect transition for graphene and graphite are compared with the
  experimental measurements for highly-oriented pyrolytic graphite and
  a single-crystal of graphite.  The dispersion relations for the
  transverse optical (TO) and the out-of-plane longitudinal acoustic
  (ZA) phonon modes of graphite and the TO phonon mode of graphene can
  be extracted from the inelastic ARPES intensity.  We find that the
  TO phonon mode for $\mathbf{k}$ points along the
  $\mathit{\Gamma}$--$K$ and $K$--$M$--$\kprime$ directions in the
  Brillouin zone can be observed in the ARPES spectra of graphite and
  graphene by using a photon energy $\approx 11.1$ eV. The relevant
  mechanism in the ARPES process for this case is the resonant
  indirect transition.  On the other hand, the ZA phonon mode of
  graphite can be observed by using a photon energy $\approx 6.3$ eV
  through a nonresonant indirect transition, while the ZA phonon mode
  of graphene within the same mechanism should not be observed.
\end{abstract}


\maketitle

\section{Introduction}
Angle-resolved photoemission spectroscopy (ARPES) is one of the
well-known methods to probe the electron-phonon interaction in
solids~\cite{damascelli03}.  Renormalization of the electronic energy
due to the electron-phonon interaction has been explored by the
observation of the electron dispersion relation near the Dirac point
(the $K$ or $K'$ points of the hexagonal Brillouin zone) in
graphene~\cite{bostwick07,calandra07,tse07a}.  The electron-phonon
renormalization causes the appearance of a kink structure in the
electron dispersion relation.  The ARPES intensity is expressed in
terms of the self-energy, in which the real and imaginary parts of the
self-energy determine the kink structure and the linewidth in the
electronic energy dispersion, respectively~\cite{mahan00a,grimvall80}.

It is known that the ARPES spectra around the $\Gamma$ point and near
the Fermi level (with binding energies around
$E_b\approx0$--$3\unitev$) do not exist for the direct optical
transition because there is no corresponding energy state
~\cite{bostwick07}.  However, recent ARPES experiments show that
measurement of the ARPES intensity around the $\mathit{\Gamma}$ point
and near the Fermi level could also provide valuable information on
the electron-phonon interaction~\cite{liu10aa,tanaka13}.  For example,
Liu \emph{et al.}  have observed the ARPES spectra at the
$\mathit{\Gamma}$ point and near the Fermi level for graphene-based
materials~\cite{liu10aa}.  They pointed out that the observation of
ARPES spectra originates from the \emph{indirect transition} of
electrons, which is mediated by phonons.  In their experiment, the
observed ARPES spectra with binding energies around $154\unitmev$ and
$67\unitmev$ have been ascribed to the energy and momentum of the
phonon at the $K$ (or $\kprime$) point.  They suggest that the
electron is scattered from the $K$ to the $\Gamma$ point by emitting a
phonon through an indirect transition.  However, the phonon dispersion
from their experiment could not be determined because they used photon
energies of more than $20\unitev$.

Tanaka \emph{et al.}~have reported ARPES spectra of highly-oriented
pyrolytic graphite (HOPG) around the $\mathit{\Gamma}$ point and near
the Fermi level for various photon energies less than
$15\unitev$~\cite{tanaka13}.  This experiment probes the energies and
momenta of the electrons and phonons involved in the indirect
transition, for different photon energies, so that the phonon
dispersion of HOPG can be obtained.  They found that, when the
incident $p$-polarized photons are incident on the sample surface, the
ARPES intensity increases like a step-function at the binding energies
around $154\unitmev$ and $67\unitmev$ for $\hbar\omega=11.1\unitev$
and for $\hbar\omega=6.3\unitev$, respectively, and that the ARPES
spectra cannot be observed for incident photons in the range
$\hbar\omega=13$--$15\unitev$ in their experiment.  However, not all
the possible phonon dispersion relations of graphite could be
well-resolved since HOPG is not a single crystal of graphite.  Thus,
the phonon modes involved in the indirect transition were not assigned
properly from previous experimental measurements which were based on
HOPG.

In this work, motivated by the observations of the indirect transition
ARPES spectra~\cite{liu10aa,tanaka13}, we investigate the detailed
mechanisms of indirect transitions in graphene and graphite by the
calculation of the electron-photon interaction and the electron-phonon
interaction based on first principles
calculations~\cite{ayria15,hamaguchi01,ridley13,giannozzi09}.  By
considering the indirect transition, we compare our theoretical
calculation of the ARPES intensity with the latest data from
experimental measurements for HOPG and a single crystal of graphite in
order to evaluate and assign the origin of the observation of the
phonon dispersion in graphene and graphite through ARPES for various
photon energies.  We find that the dispersion relations of the
transverse optical (TO) and of the out-of-plane acoustic (ZA) phonon
modes of graphene and graphite which have even symmetry with respect
to a mirror plane (i.e., a plane which includes the incident light and
the ejected photoemission electron) can be extracted from the
experimental ARPES intensity.  Although the longitudinal acoustic (LA)
phonon mode also has the even symmetry with respect to the mirror
plane, the LA phonon mode cannot be observed due to the negligibly
small electron-phonon matrix element in the vicinity of the $K$ point.

We also find that the ARPES spectra near the binding energy of
$154\unitmev$ can be assigned by the ARPES intensity calculation as
the TO phonon mode for the applied photon energy $\hbar\omega \approx
11.1\unitev$, in which the relevant mechanism involves the resonant
indirect transition.  On the other hand, for the lower photon energy
$\hbar\omega \approx 6.3\unitev$, the ZA phonon mode is assigned to
the ARPES spectra at the binding energy of $67\unitmev$ through the
nonresonant indirect transition with $p$-polarized light.  In the case
of the photon energy $\hbar\omega=13-15\unitev$, we will show that the
ARPES intensity cannot be observed because the directions of the
ejected electron and the detector are not properly aligned with
respect to each other to have a significant ARPES intensity.
 
This paper is organized as follows.  In Sec.~\ref{sec:method} we
describe the geometry of the ARPES measurements, the experimental
setup, and the theoretical formulation of the indirect transition.  We
also discuss the symmetry considerations for the ARPES spectra by
group theory.  In Sec.~\ref{sec:results}, theoretical results for the
ARPES spectra are compared with the experimental measurements.
Finally, the summary of this paper is given in Sec.~IV.

\begin{figure}
\includegraphics[width=8cm]{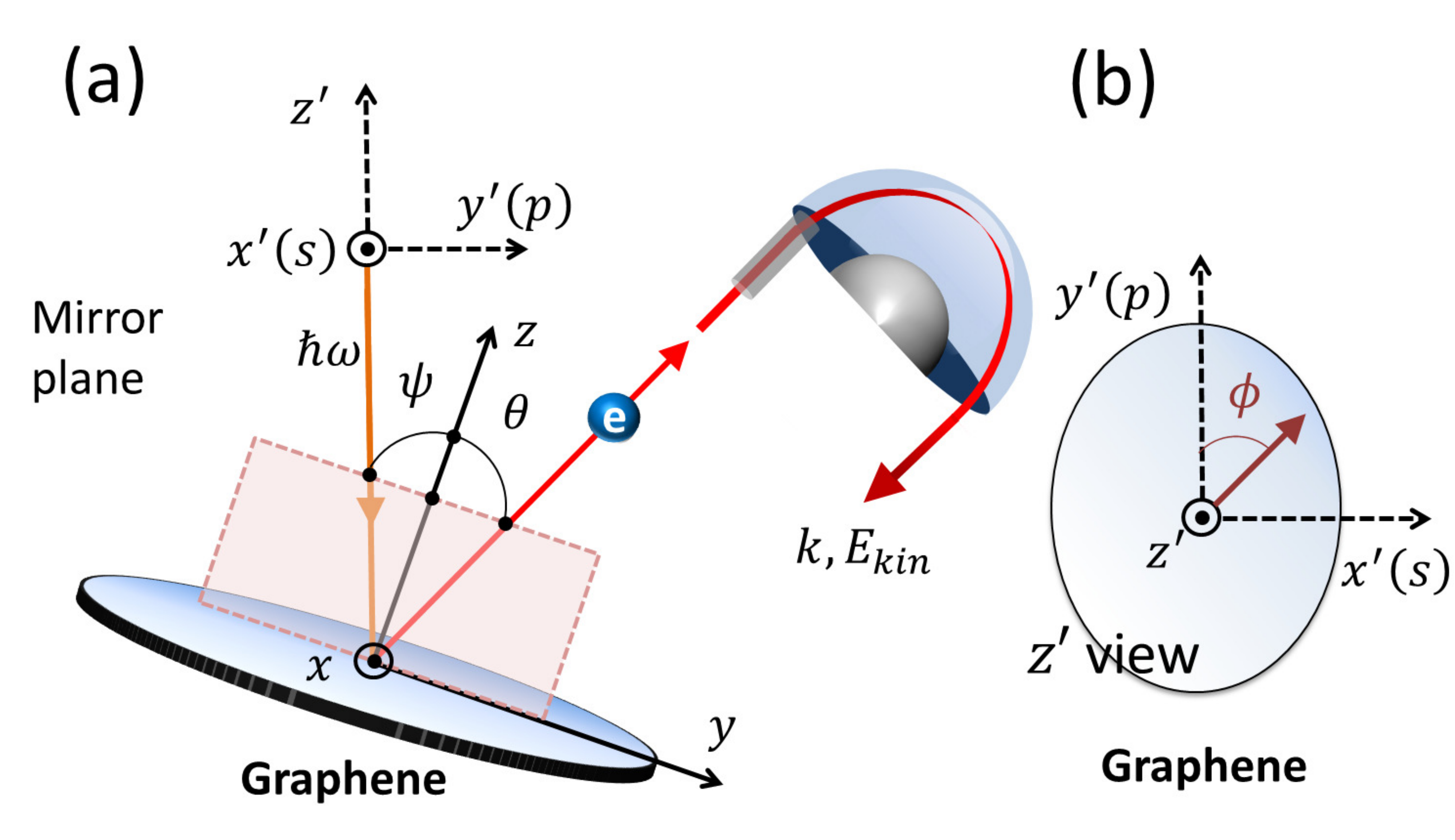}
\caption{\label{Fig.1}(Color online)(a) Geometry of the photoemission
  process~\cite{ayria15}.  The incident photon with energy
  $\hbar\omega$ is shown by an arrow going to the graphene plane.  We
  can define a mirror plane which contains the directions of the
  incident light ($z^\prime$ axis), the electrons ejected from the
  surface, and an axis ($z$-axis) normal to the graphene surface.  The
  angles between the incident light, the ejected electron, and the
  $z$-axis is denoted by $\psi$, $\theta$. (b) Viewing the set-up from
  the $z^\prime$ axis, the light polarization angle, $\phi$, is in the
  $x^\prime y^\prime$-plane and is measured with respect to the
  $y^\prime$ axis. Here, $\phi=0^\circ$ and $\phi=90^{\circ}$
  correspond to the $p$- and $s$-polarizations, respectively.}
\end{figure}

\section{Methods}
\label{sec:method}

\subsection{ARPES experiments}
\label{sec:exp}
In Fig.~\ref{Fig.1}, the experimental setup is shown schematically, in
which the graphene surface is irradiated by photons having an incident
angle $\psi$ with respect to the $z$-axis, normal to the surface.  The
emitted electrons with an emission angle $\theta$ are analyzed with
respect to the kinetic energy and momentum~\cite{Luth95}.  When we
look at the surface in the direction of the $z'$ axis, we see that the
light polarization angle $\phi$ is defined in the $x'y'$-plane and
measured by the $y'$-axis, as shown in Fig.~\ref{Fig.1}(b).  Here,
$\phi=0^{\circ}$ and $\phi=90^{\circ}$ correspond to the $p$- and
$s$-polarization directions,
respectively~\cite{ayria15,kruczynski08,liu11l,hwang11b}.  In this
paper, we adopt a particular geometry so that the analyzer is
perpendicular to the surface, i.e., $\theta=0$.  The experiments were
carried out at two beamlines of the synchrotron radiation facilities:
(1) BL-7U facility of UVSOR at the Institute for Molecular Science,
Okazaki, Japan and (2) BL-9A facility of HiSOR at Hiroshima
University, Higashi-Hiroshima, Japan.  The photon energy dependence of
the HOPG was taken at the BL-7U facility and the
polarization-dependence of the single-crystalline graphite was taken
at the BL-9A facility.  In both beamlines, the combination of the
APPLE-II-type variable-polarization undulator, normal incidence
monochromator, novel photoelectron spectrometer with a multichannel
detector, and a precise multiaxis goniomer for the sample enables us
to measure the ARPES spectra with sufficient resolution for observing
the phonons ($<10\unitmev$).  The sample was kept at $30\unitkelvin$
during the measurement with a He-cryostat.

\subsection{Formulation of ARPES intensity}

Let us define the Hamiltonian; $H_e$ for electrons, $H_{ph}$ for
phonons, $H_{\rm{opt}}$ for the electron-photon interaction and
$H_{\rm{ep}}$ for the electron-phonon interaction.  The total
Hamiltonian, $H$, is written as
\begin{equation}
  H = H_e + H_{p} + H_{\rm{opt}}+H_{\rm{ep}},
\label{eq:tham}
\end{equation}  
where the unperturbed Hamiltonian of electrons and phonons is
considered as $H_0 = H_e+H_{p}$.  We adopt the adiabatic
approximation, which implies that the total wave function can be
written as a product of an electron eigenstate and phonon
eigenstate~\cite{grimvall80}.

The unperturbed electron and phonon dispersion relation and their
eigenstates for points along the high symmetry axes are calculated
using the \texttt{Quantum ESPRESSO} package~\cite{giannozzi09}.  For
the electron calculation, we adopt the norm-conserving pseudopotential
with the Perdew-Zunger exchange-correlation scalar relativistic
functional.  The kinetic energy cut-off is taken as $60\unitry$ for
each atom and the kinetic energy cut-off for the electron density
potential is set to be $600\unitry$ in order to verify the convergence
of all wave functions.  The $k$-points for self-consistent calculation
are taken within the $42\times 42\times 1$ and $20\times 20\times 4$
mesh grids in the graphene and graphite Brillouin zones, respectively.
The lattice parameter of graphene is considered to be
$2.46\unitangstrom$ and the lattice constant in the direction normal
to the graphene plane is taken as $c/a= 20.0$ and $c/a= 2.7$ for
graphene and graphite, respectively.  As for the phonon calculation,
we adopt the Perdew-Burke-Ernzerhof generalized gradient approximation
for the exchange-correlation function~\cite{perdew96}.  The kinetic
energy cut-off is set to be $100\unitry$ for each atom, while the
kinetic energy cut-off for the density potential is taken as
$1200\unitry$.  Following Ref.~\onlinecite{mounet05}, the dynamical
matrix is calculated on $6\times 6\times 1 $ and $6\times 6\times 4 $
$q$-point mesh grids in graphene and graphite, respectively, where $q$
is the phonon wavevector.

\begin{figure}
\includegraphics[width=8cm]{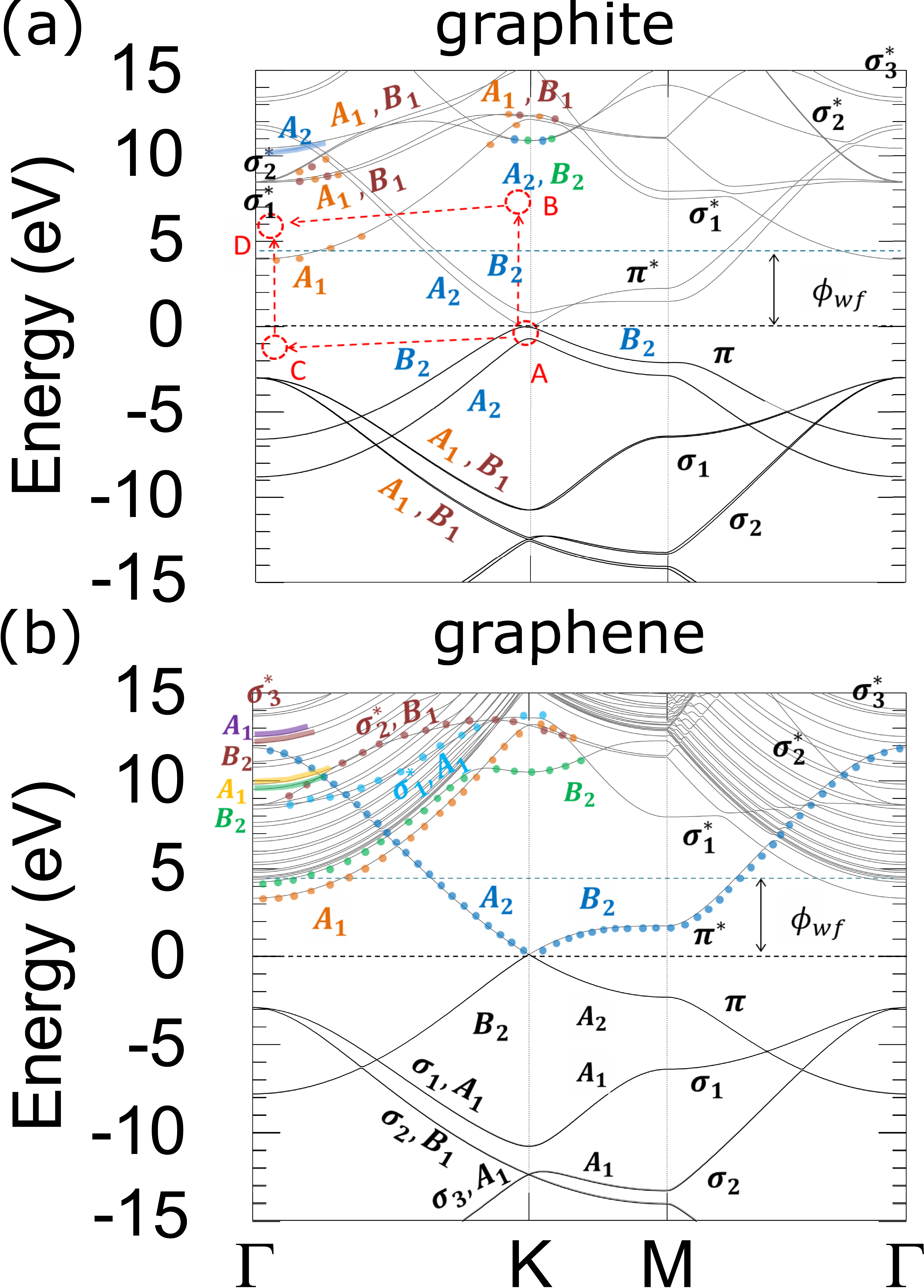}
\caption{\label{Fig.2}(Color online) Electronic energy dispersion
  relations of (a) graphite and (b) graphene are calculated by
  first-principles calculations and plotted along the high symmetry
  directions $\mathit{\Gamma}$--$K$--$M$--$\mathit{\Gamma}$ up to
  $15\unitev$.  In panel (a), the two possibilities of indirect
  transitions (A $\rightarrow$ B $\rightarrow$ D and A $\rightarrow$ C
  $\rightarrow$ D) are shown by the red dash-dotted arrows, in which
  an electron from the initial state A can reach the final state D
  mediated by electron-phonon interaction.  The separation between the
  states A and B (or C and D) is determined by the incident photon
  energy used in ARPES (in this picture it is $\sim 7\unitev$).  Note
  that in both panels (a) and (b) we show some symmetry
  representations for the energy bands which might be involved in the
  indirect transitions in (a) graphite and (b) graphene.}
\end{figure}

\begin{figure}
\includegraphics[width=8cm]{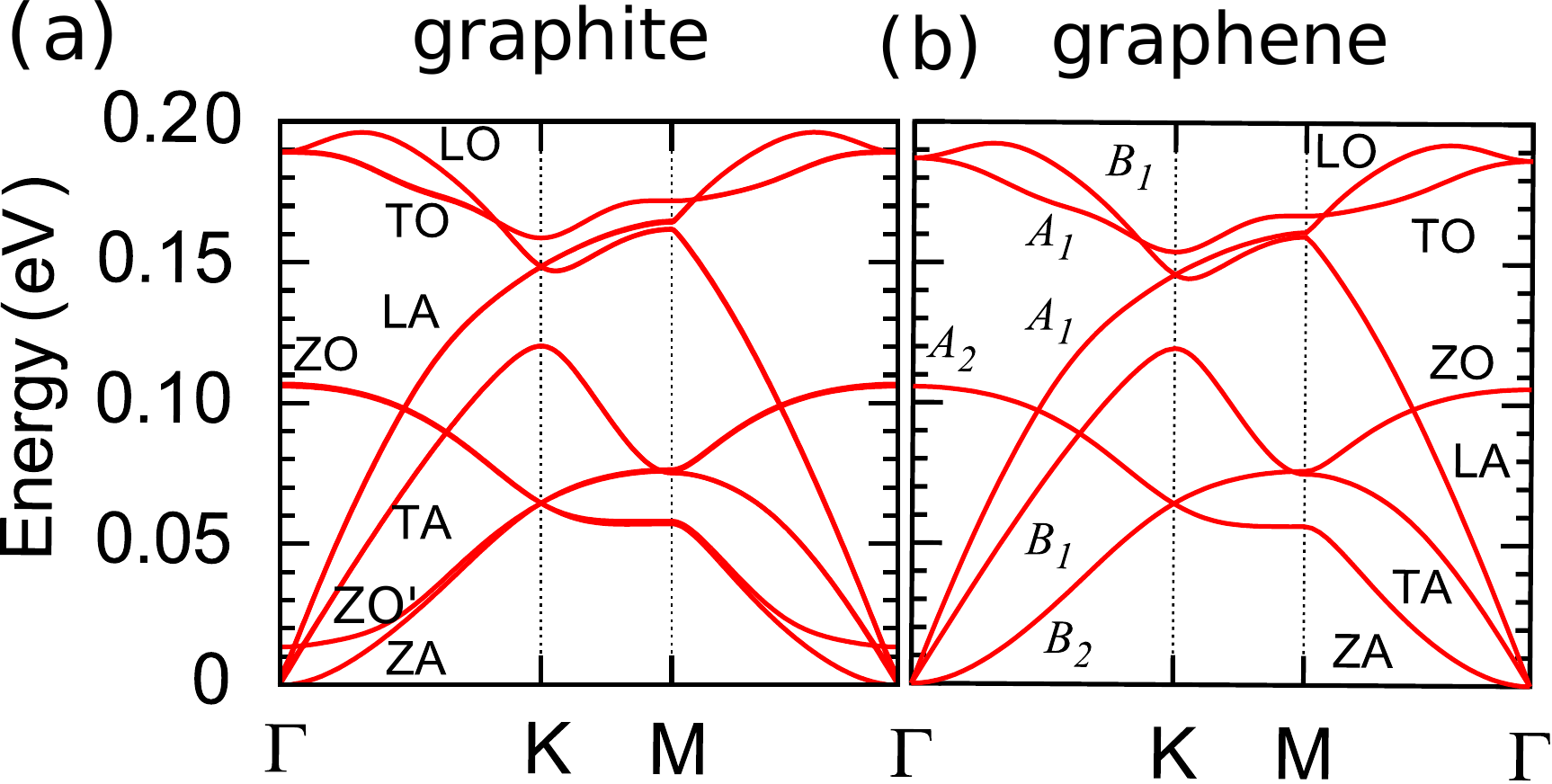}
\caption{\label{Fig.3}(Color online) The phonon energy dispersion
  relations for (a) graphite and (b) graphene, obtained from
  first-principles calculations and density functional perturbation
  theory~\cite{giannozzi09}.  Since there is $C_{2v}$ symmetry along
  the $\mathit{\Gamma}$--$K$--$M$ and $\mathit{\Gamma}$--$M$
  directions, each phonon mode is labeled by the irreducible
  representation of the $C_{2v}$ point group along the
  $\mathit{\Gamma}$--$K$--$M$ direction~\cite{reich04}. The TA, LA TO
  and LO phonon modes along the $\Gamma$--$K$--$M$ correspond to
  $B_1$, $A_1$, $A_1$ and $B_1$ symmetries, respectively.  The ZA and
  ZO phonon modes along the $\Gamma$--$K$--$M$ correspond to $B_2$ and
  $A_2$ symmetries.  }
\end{figure}

The calculated electron energy dispersions of graphite and graphene
are shown in Figs.~\ref{Fig.2}(a) and (b), respectively, whereas the
calculated phonon dispersion of graphite and graphene are shown in
Figs.~\ref{Fig.3}(a) and (b) respectively.  Since there are four atoms
in the unit cell of graphite, there will be twelve phonon modes.  Most
of the phonon modes are nearly doubly degenerate and similar to those
for graphene~\cite{dresselhaus77l,mohr07}.  Thus, we put the same
labels of phonon modes for graphite and graphene, except for the
breathing mode ($\text{ZO}^\prime$) of two layers in graphite.  In
graphene, there are six phonon modes which consist of four in-plane
modes and two out-of-plane modes.  At the $\Gamma$ point, there are
three acoustic branches: (1) the transverse acoustic, (2) the
longitudinal acoustic, and (3) the out-of-plane acoustic phonon modes,
which are labeled in Fig.~\ref{Fig.3}(b) as TA, LA, and ZA,
respectively.  Unlike the case of graphite, there is no ZO$^\prime$
mode in graphene.  There are also three optical phonon modes in
graphene above $0.1\unitev$ at the $\Gamma$ point: (1) the transverse
optical, (2) the longitudinal optical, and (3) the out-of-plane
optical phonon mode, which are labeled as TO, LO, and ZO,
respectively. The symmetry labels of phonon modes of graphene are also
shown in Fig.~\ref{Fig.3}(b) which will be discussed in the next
section.

The perturbation Hamiltonian in Eq.~\eqref{eq:tham} is considered as
\begin{equation}
  H'=H_{\rm{opt}}+H_{\rm{ep}}.
\end{equation}  
The transition rate from an initial state $|i\rangle$ to a final state
$|f\rangle$ through a virtual state $|m\rangle$ is given by the
second-order time-dependent perturbation theory
\cite{hamaguchi01,ridley13},
\begin{equation}
 \begin{split}
 W(\mathbf{k}_f,\mathbf{k}_i) & =  
 \frac{2\pi}{\hbar} \bigg\vert S(\mathbf{k}_f,\mathbf{k}_i) \bigg\vert  ^2 
\delta(\varepsilon_i-\varepsilon_f),
 \end{split}
\label{eq:indirect}
\end{equation}
where $\varepsilon_i$ and $\varepsilon_f$ represent the energy of an
initial state and a final state, respectively, and the matrix
$S(\mathbf{k}_f,\mathbf{k}_i)$ is given by
\begin{equation}
 \begin{split}
 S(\mathbf{k}_f,\mathbf{k}_i) & =  
  \sum_m\frac{\langle f |H'
 |m  \rangle \langle m |H'|i \rangle}{\varepsilon_i-\varepsilon_m}.
 \end{split}
\label{eq:indirect1}
\end{equation}

There are two scattering processes following Eq.~\eqref{eq:indirect1}
that may contribute to the indirect transition, i.e.,
A~$\rightarrow$~B~$\rightarrow$~D and
A~$\rightarrow$~C~$\rightarrow$~D, which are depicted as red
dash-dotted arrows in Fig.~\ref{Fig.2}(a).  In the
A~$\rightarrow$~B~$\rightarrow$~D process: (1) a photon excites an
electron from the initial state $|A\mathbf{k}_i\rangle$ to a state
$|B\mathbf{k}_m\rangle$ and then (2) the photoexcited electron from
the state $|B\mathbf{k}_m\rangle$ is scattered to the final state
$|D\mathbf{k}_f\rangle$ by emitting a phonon.  Since the temperature
of the sample is considered to be $60\unitkelvin$, the probability
absorption of a phonon in the A~$\rightarrow$~B~$\rightarrow$~D
process is negligible.  In the A~$\rightarrow$~C~$\rightarrow$~D
process: (1) a phonon scatters an electron from the initial state
$|A\mathbf{k}_i\rangle$ to a state $|C\mathbf{k}_m\rangle$ and then
(2) a photon excites the scattered electron from the state
$|C\mathbf{k}_m\rangle$ to the final state $|D\mathbf{k}_f\rangle$.
All these processes are expressed by the following equation:
\begin{align}
  S(\mathbf{k}_f,\mathbf{k}_i) &= \frac{\langle D\textbf{k}_f,
    |H_{\rm{ep}} |B\textbf{k}_m \rangle \langle B\textbf{k}_m
    |H_{\rm{opt}}|A\textbf{k}_i \rangle}{
    E_i(\mathbf{k}_i)+\hbar\omega - E_B(\mathbf{k}_i)}
  \notag\\
  &+ \frac{\langle D\textbf{k}_f |H_{\rm{opt}} | C\textbf{k}_m \rangle
    \langle C\textbf{k}_m |H_{\rm{ep}}|A\textbf{k}_i
    \rangle}{E_i(\mathbf{k}_i) - \hbar\omega_q^\alpha-
    E_C(\mathbf{k}_f)},
\label{eq:extend}
\end {align}                   
where the energy and momentum conservation requirement gives the
energy denominators of Eq.~\eqref{eq:extend} and $\mathbf{k}_f
=\mathbf{k}_i + \mathbf{q}$.  In Eq.~\eqref{eq:extend}, $\hbar\omega$
denotes the photon energy and $ \hbar\omega_q^\alpha$ refers to the
energy of the $\alpha$th phonon mode with the wave vector
$\mathbf{q}$.  After calculations, we find that the
A~$\rightarrow$~B~$\rightarrow$~D transition would be more dominant
than the A~$\rightarrow$~C~$\rightarrow$~D transition both for photon
energies around $\hbar\omega\approx 11\unitev$--$15\unitev$ and for
lower photon energies around $\hbar\omega\approx 6\unitev$.  However,
it should be noted that there are different physical origins for why
the A~$\rightarrow$~B~$\rightarrow$~D transition is always dominant in
the two different energy ranges, as will be discussed in
Sec.~\ref{sec:results}.

We adopt the rigid-ion approximation for the electron-phonon matrix
element $\langle f, \textbf{k}_f|H_{\rm{ep}} |i, \textbf{k}_i
\rangle$~\cite{grimvall80,glembocki82}, whose detailed derivation is
given in Appendix~\ref{sec:epc}.  In the case of the electron-photon
transition from an initial state $|i, \mathbf{k} \rangle$ to a final
state $|f, \mathbf{k} \rangle$, the electron-photon matrix element is
given in the dipole approximation~\cite{gruneis03}, while $\langle f,
\textbf{k}|H_{\rm{opt}} |i, \textbf{k} \rangle
\propto\mathbf{A}\cdot\mathbf{D}(\mathbf{k})$, where $\mathbf{A}$ is
the vector potential and $\mathbf{D}(\mathbf{k})=\langle f,
\textbf{k}| \mathbf{\nabla}|i, \textbf{k} \rangle$ is the dipole
vector.  The electron-photon interaction for different photon energies
based on the plane wave expansion is discussed in the previous study
\cite{ayria15}, and we follow the same method in the present work to
calculate the electron-photon matrix element.
 
To obtain the ARPES intensity $I(E,\hbar\omega)$, we need to integrate
over all the initial states and all the final states.  The summation
on the initial states and the final states can be performed
independently when we adopt the experimental conditions that are
chosen for each ARPES experiment~\cite{hufner96}.  The ARPES intensity
$I(E,\hbar\omega)$ is given by
\begin{align}
  I(E,\hbar\omega) \propto& \sum_{i,f}\int d\mathbf{k}_i d\mathbf{k}_f
  |S(\mathbf{k}_f,\mathbf{k}_i)|^2 \delta(\varepsilon_i-\varepsilon_f)
  \notag\\
  &\times \delta(E-\varepsilon_f+\phi_{\mathrm{wf}})(N_q^\alpha+1)f^{\mathrm{occ}}_F,
\label{eq:intensity}
\end{align}
where $\phi_{\mathrm{wf}}=4.5$ eV is the work function of
graphene~\cite{giovannetti08}, $f^{\mathrm{occ}}_F$ denotes the
Fermi-Dirac distribution function for the occupied state and
$N_q^\alpha$ is the quantum number of the phonon mode $\alpha$ with
wave vector $\mathbf{q}$.  The first delta function in
Eq.~\eqref{eq:intensity} expresses conservation of total energy, while
the second delta function ensures that the photoelectrons have higher
energies than the work function $\phi_{\mathrm{wf}}$.  In addition,
there are several symmetry selection rules for the ARPES spectra in
graphite and graphene, especially for the indirect transition.  These
selection rules are discussed in the next section.

\subsection{Symmetry selection rules for ARPES}
\label{sec:sym}

\begin{figure}
  \centering\includegraphics[width=5cm]{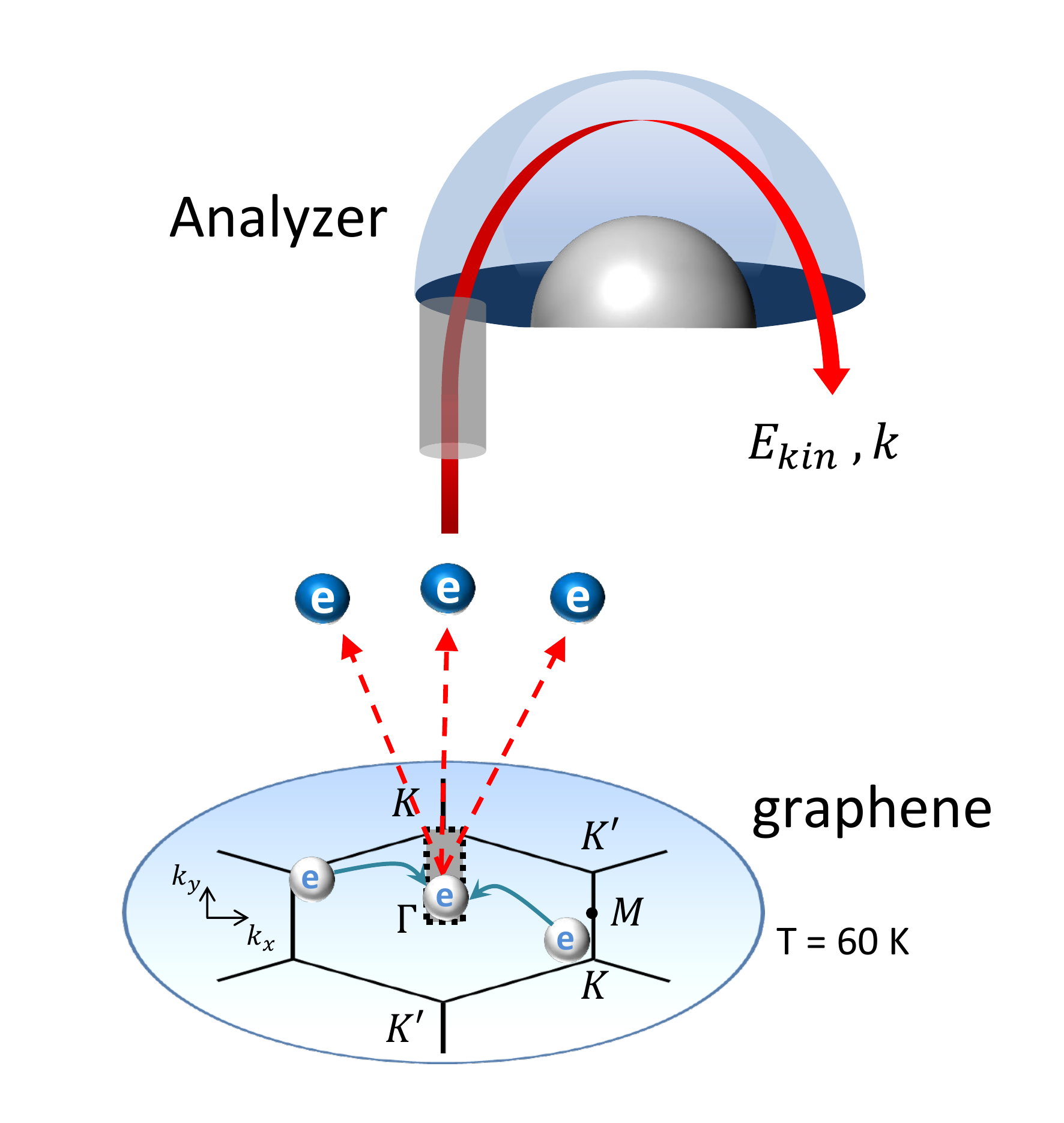}
  \caption{\label{symm}(Color online) The indirect scattering of an
    electron.  An electron around the $K$ or $K'$ point scatters
    around the $\Gamma$ point by the indirect transition.  The shaded
    region line along $\Gamma$--$K$ is what we observed for
    $E_{\text{kin}}$ and $\mathbf{k}$.}
 \end{figure}

\begin{table}[t]
  \caption {Character table of the $C_{2v}(2mm)$ point group.}
\label{tab:chartable} 
\begin{tabular}{ r  r r c c r r  c }
  \toprule
  & $E$ & {  } $C_2$ & { } $\sigma_v(xz)$ & { } { }{ }$\sigma'_v(yz)$
  &{   }{  }{  }bases&   \\
  \hline
  $A_1$& 1   &  1    &  {}{    } 1        & {}{    } 1         &{  } {
  }$z,$       $\nabla_z$ \\ 

  $A_2$& 1   &  1    &  $-1$      & $-1$       &{  } {  }$R_z$     \\ 
  $B_1$& 1   & $-1$  &  {}{    } 1        & $-1$       &{  } {
  } $x,R_y,$   $\nabla_x$\\ 

  $B_2$& 1   & $-1$  &  $-1$      &  { }{    }1         &{  } {
  }$y,R_x,$   $\nabla_y$\\ 
  \toprule
\end{tabular}
\end{table}

\begin{table}[t]
  \caption {Direct product table of the $C_{2v}$ representation for the 
    indirect transition $A\rightarrow B \rightarrow D$.  Here
    $\Gamma_i$ indicates the initial states,
    $\Gamma_i=\{A_2 ,B_2\}$, while
    $\Gamma_o=\{A_1, B_1, B_2\}$ refers to the optical transition
    and $\Gamma_q$ assigns the phonon eigenvector 
    symmetry along $\Gamma$--$K$--$M$.
    For the final states $\Gamma_f=\{A_1 ,B_2\}$, the corresponding column 
    shows the symmetry of the allowed final state.}
\label{tab:chartable2} 
\begin{tabular}{ c c c | c c c c c  }
 \toprule
 {	}$\Gamma_i${	}& {    }$\Gamma_o$(Pol.){    }&{    }
$\Gamma_m${    }&{    }$\Gamma_m${    }&{    }
$\Gamma_q$(Ph.){    }&{   }$\Gamma_f${    }&&   \\
 \hline
 $B_2$& $A_1$($p$)   &  $B_2 $   &  $B_2 $   &  $A_1$(TO,LA)      &    $ B_2$     &\\ 
 $B_2$& $B_2$($p$)   &  $A_1 $   &  $A_1 $   &  $A_1$(TO,LA)      &    $ A_1$     &\\ 
 $B_2$& $A_1$($p$)   &  $B_2 $   &  $B_2 $   &  $B_2$(ZA)        &    $ A_1$     &\\ 
 $B_2$& $B_2$($p$)   &  $A_1 $   &  $A_1 $    &  $B_2$(ZA)      &    $ B_2$     &\\ 

 $A_2$& $B_1$($s$)   &  $B_2 $   &  $B_2 $   &  $A_1$(TO,LA)      &    $ B_2$     &\\ 
 $A_2$& $B_1$($s$)   &  $B_2 $   &  $B_2 $   &  $B_2$(ZA)        &    $ A_1$     &\\ 

 \toprule
\end{tabular}
\end{table}

The geometry of the indirect scattering of electrons is schematically
illustrated in Fig.~\ref{symm}.  The electrons around the $K$ or
$\kprime$ point can scatter to the region near the $\Gamma$ point by
an indirect transition.  The shaded region along $\Gamma$--$K$
direction displays the locations where the photoemitted electrons are
measured in the ARPES experiment.  The measurement of the ARPES
intensity along the $\mathit{\Gamma}$--$K$ line provides information
about the phonon dispersion along $\mathit{\Gamma}$--$\kprime$ and
$K$--$M$--$\kprime$~\cite{liu10aa}, which is explained in
Appendix~\ref{sec:lsym}.

In graphene and graphite, the three high-symmetry points $\Gamma$,
$K$(or $K'$), and $M$ correspond to the $D_{6h}$, $D_{3h}$ and
$D_{2h}$ point group symmetries, respectively.  The electronic states
along the $K'$--$\Gamma$--$K$ and $K'$--$M$--$K$ lines belong to the
$C_{2v}$ point group, while any other general $\mathbf{k}$ points
belong to the $C_{1h}$ point group~\cite{malard09g,kogan12,kogan14}.
The $C_{2v}$ group has four irreducible representations
$\{A_1,A_2,B_1,B_2\}$ as shown in Table I.  According to the $x,y,z$
coordinates in Fig.~\ref{symm}, the character Table of $C_{2v}$ is
listed in Table I.  Moreover, in the $C_{2v}$ symmetry, there are two
mirror plane operations $\sigma_v(xz)$ and $\sigma'_v(yz)$ in the
Brillouin zone.  The plane $\sigma_v(xz)$ is aligned with the
$M$--$\Gamma$--$M$ line, while the plane $\sigma'_v(yz)$ is aligned
with the $K$--$\Gamma$--$K'$ line.  Since we observe ARPES spectra
in the present study, along the $K$--$\Gamma$ lines, $\sigma'_v(yz)$ is
relevant to the ARPES spectra as shown below.  As we explained, the
symmetry labels of each of the phonon modes of graphene along the
$\Gamma$--$K$--$M$ lines are shown in Fig.~3(b).

The selection rules for the optical and lattice vibrations impose
nonzero matrix elements $\langle m |o|i\rangle$ and
$\langle f |q|m\rangle$ in Eq.~(\ref{eq:extend}), which
satisfy ~\cite{dresselhausb08}:
\begin{equation} 
\begin{split} 
\Gamma_m \otimes \Gamma_o \otimes \Gamma_i = A_1\quad \text{and} \quad
\Gamma_f \otimes \Gamma_q \otimes \Gamma_m = A_1
\end{split} 
\label{eq:sym2} 
\end {equation}
where $\Gamma_i$, $\Gamma_m$, $\Gamma_f$, $\Gamma_o$, $\Gamma_q$ are,
respectively, the irreducible representations for the initial state,
intermediate state, and final state of the electron wave functions, the
dipole vector and the phonon mode.

Further, in the ARPES experiment, we need also to consider the
conservation of parity for the matrix elements of the product $\langle
f|H'|m\rangle \langle m|H'|i\rangle$, in Eq.~(\ref{eq:indirect1}),
under reflection from the mirror plane $\sigma'_v(yz)$~\cite{Luth95}.
This condition imposes an additional restriction to get a non-zero
ARPES intensity, i.e., the integral of $\langle f|H'|m\rangle \langle
m|H'|i\rangle$ must be an even function for the $\sigma'_v(yz)$
symmetry operation.  Furthermore, it is known that the final state,
$\langle f|$ must have even symmetry with respect to the mirror plane
$\sigma'_v(yz)$ in the ARPES
experiment~\cite{Luth95,damascelli03,mahatha12,liu11l}.

If we consider the $\pi$ band as the initial state, $\Gamma_i$,
this state has to have $B_2$ and $A_2$ symmetry, as shown in Fig.~\ref{Fig.2},
 \begin{equation} 
 \begin{split} 
\Gamma_i=A_2\quad\text{or}\quad B_2\notag.
\end{split} 
\label{eq:initial} 
\end {equation}
Since the final state, $\Gamma_f$ has to have even symmetry with respect to the
mirror plane, $\sigma'_v(yz)$ , the final state belongs to the $A_1$ or $B_2$
irreducible representation, 
 \begin{equation} 
 \begin{split} 
\Gamma_f=A_1\quad\text{or}\quad B_2\notag.
\end{split} 
\label{eq:initial} 
\end {equation}

Since most of the phonon branches of graphite are nearly doubly
degenerate and almost similar to those in graphene, in the following
symmetry discussion, we will use the symmetry of the phonons in
graphene for simplicity~\cite{mohr07}.  Graphene has six phonon modes,
as shown in Fig.~\ref{Fig.3}(b). The four in-plane phonon modes TA,
LA, TO and LO along $\Gamma$--$K$--$M$ transform as $B_1$, $A_1$,
$A_1$ and $B_1$, respectively.  The-out-of-plane phonon modes ZA and
ZO along $\Gamma$--$K$--$M$ transform as $B_2$ and $A_2$, respectively
\cite{thomsen07,maultzsch04,mohr07}.  Moreover, the parity under the
$\sigma'_v(yz)$ reflection along the $\Gamma$--$K$ line is odd for TA,
LO, ZO and is even for ZA and LA, TO~\cite{juan15}. Thus, only the ZA,
LA and TO phonon modes can be observed in the ARPES spectra.  It is
important to note that the $p$-polarized light lies in the yz-plane,
shown in Fig.~\ref{Fig.1}, and it transforms as the $B_2$, $A_1$
irreducible representations.  The $s$-polarized light is parallel to
the $x$-axis and it transforms as the $B_1$ irreducible
representation.  In order to satisfy Eq.~\eqref{eq:indirect1} there
are six possibilities in the ARPES processes for $\langle m
|o|i\rangle$ and $\langle f |q|m\rangle$, which are summarized in
Table II.

\section{Results and discussion}
\label{sec:results}

\begin{figure}
\includegraphics[width=80mm]{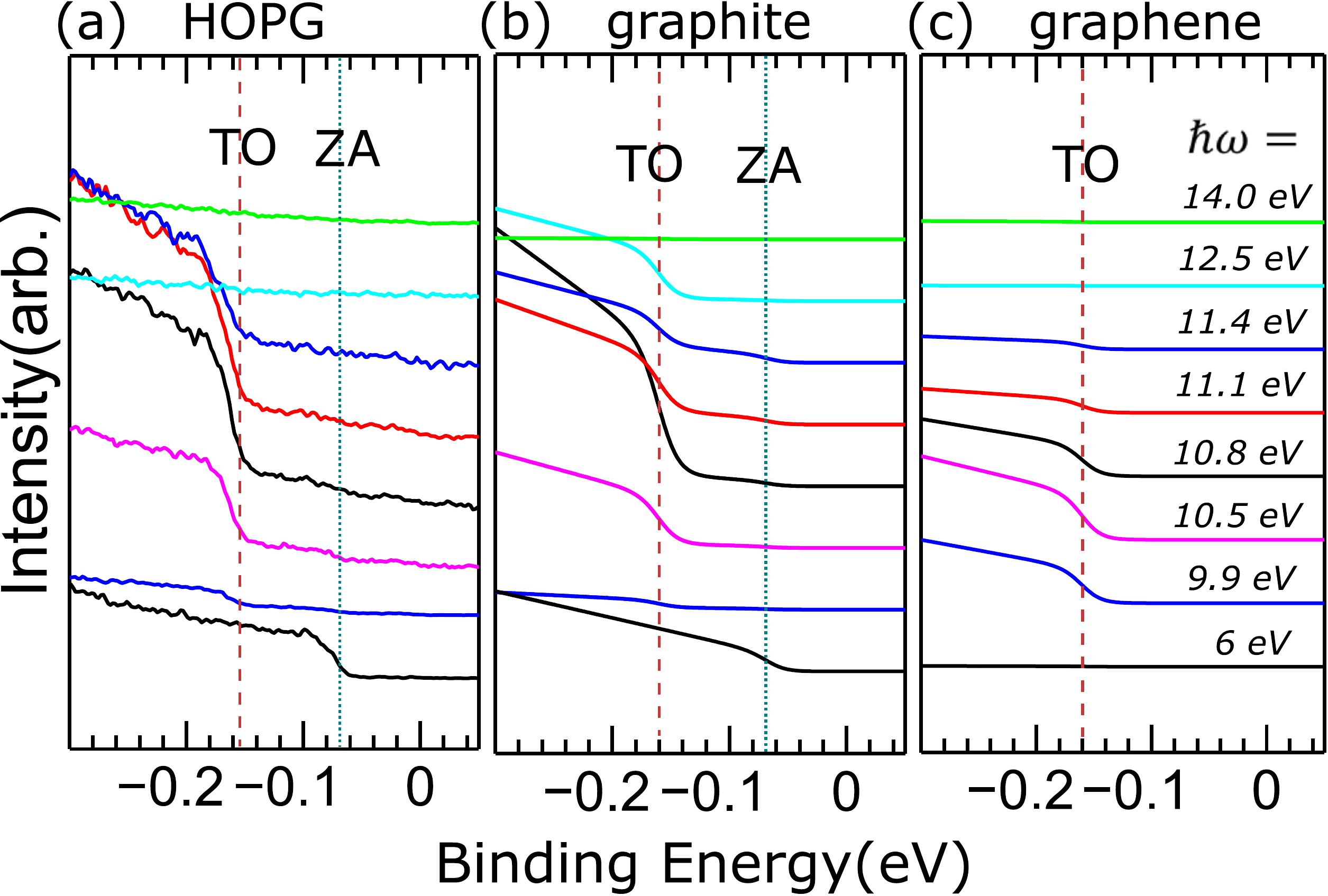}
\caption{\label{exth} (Color online) (a) The experimental ARPES
  intensities for HOPG compared with the calculated ARPES intensities
  for (b) graphite and (c) graphene near the $\Gamma$ point for
  several incident photon energies.  The incident photon is
  $p$-polarized light and the incident angle is $\psi=40^{\circ}$.  In
  (a), step-like features are found at the binding energy $E_b \approx
  154\unitmev$ (dashed line) and $E_b\approx67\unitmev$ (dotted line), which
  are assigned to the TO and ZA modes, respectively. In (b), the
  step-like features from the calculations for the TO and ZA modes are
  found to be at $E_b \approx 160\unitmev$ and $E_b\approx67\unitmev$,
  respectively. In (c), from our calculation, we find only the TO
  mode, but no ZA mode.}
\end{figure}

To investigate the observation of the ARPES spectra at the $\Gamma$
point and near the Fermi level energy, we calculate here the indirect
transition ARPES intensity as a function of the binding energy, for
the $\mathbf{k}$ vectors very close to the $\Gamma$ point, at
$\mathbf{k}=2\pi \slash a\times 10^{-4}$, along $\Gamma$--$K$ for
several photon energies of graphene and graphite.  Here, we consider
the $p$-polarized light with an incident angle $\psi=40^\circ$ that is
the same as the experimental setup.  The, calculated results can then
be compared with the experimental ARPES spectra.

Figure~\ref{exth}(a) shows the experimental ARPES intensity as a
function of the binding energy for highly-oriented pyrolytic graphite
(HOPG), whereas Figs.~\ref{exth}(b) and (c) show the calculated 
ARPES intensity for graphite and graphene, respectively.  Looking at
Figs.~\ref{exth}(a) and (b), the calculated ARPES intensity basically
reproduces the experimental data.  We can see that there are step-like
features in the ARPES intensity at the binding energies
$E_b\approx154\unitmev$ and $E_b\approx67\unitmev$ for the
experimental measurements and at $E_b \approx160 \unitmev$ and $E_b
\approx 67\unitmev$ for the corresponding theoretical calculations for
graphite.  The small discrepancy between the experiment and theory for
the positions of the step-like features might originate from the Kohn
anomaly~\cite{mahan00a}, which is neglected in our calculations for
simplicity.  We assign the step-like features at
$E_b\approx154\unitmev$ (or $160\unitmev$) and at $E_b \approx
67\unitmev$ to the TO and ZA modes, respectively.  Furthermore, in
Fig.~\ref{exth}(c), we can see the step-like features only at
$E_b\approx160\unitmev$ and there is no such feature at $E_b \approx
67\unitmev$.  In the present work, we only perform the calculations
for graphite and monolayer graphene.  However, for the TO mode, we
expect that the ARPES intensity in the case of few-layer graphene
might show similar results to that of graphite.  As for the ZA mode,
few-layer graphene might show a transition from the feature of
monolayer graphene to graphite.  We will understand all these
behaviors by discussing the detailed scattering processes in the
following subsections.

\subsection{Resonant indirect transitions}

\begin{figure}
\includegraphics[width=65mm]{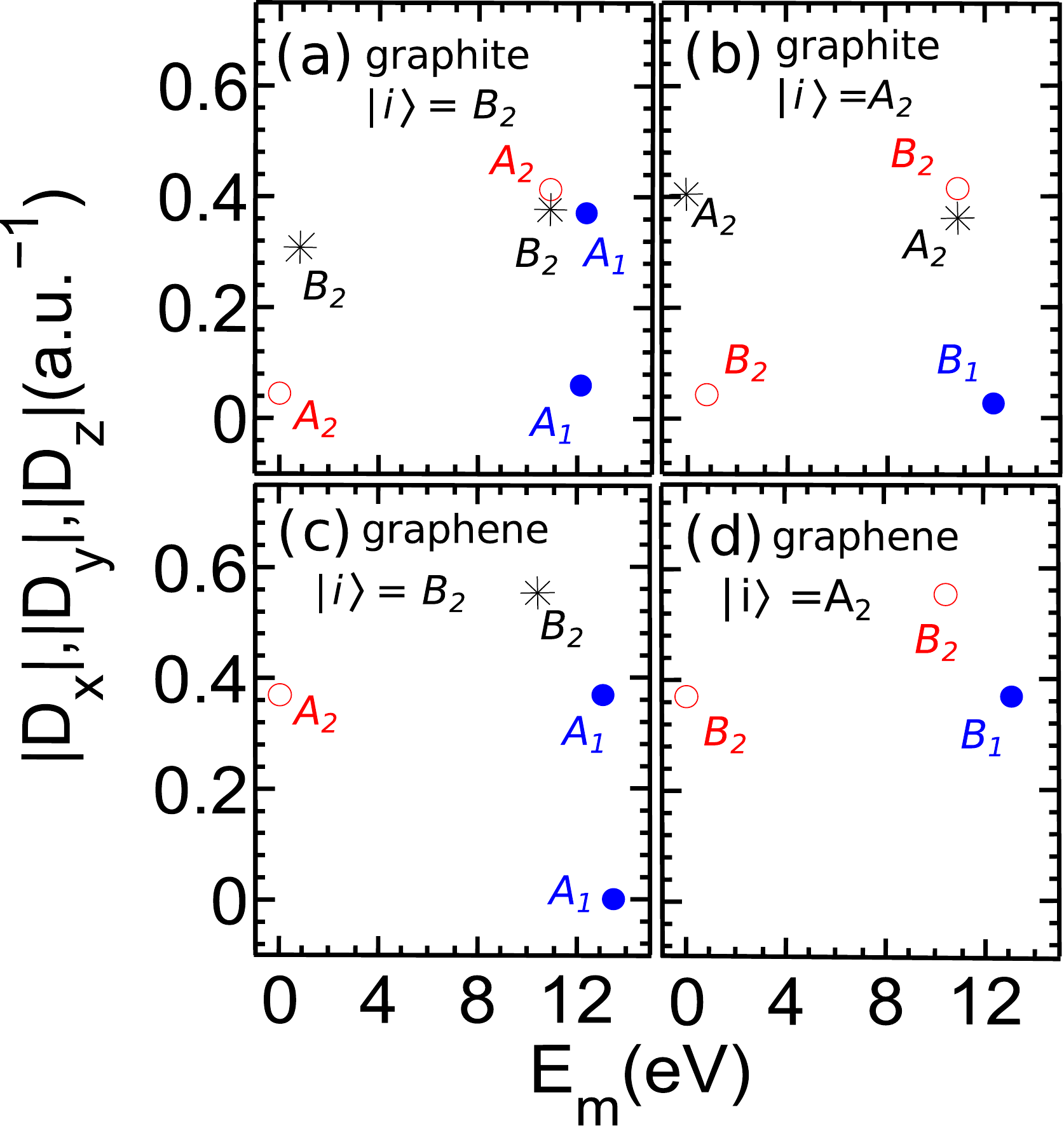}
\caption{\label{dip} (Color online) The $x$-, $y$-, and $z$-
  components of the dipole vector, i.e., $D_x$ (circles), $D_y$
  (full circles), $D_z$ (asterisks), plotted as a function of the energy of
  the intermediate state ($E_m$) for (a) graphite with $B_2$ symmetry
  as the initial state, (b) graphite with $A_2$ symmetry as the
  initial state, (c) graphene with $B_2$ symmetry as the initial
  state, and (d) graphene with $A_2$ symmetry as the initial state.
  Symmetry labels near the circles, dots, and asterisks correspond to
  the symmetry of the intermediate states.}
\end{figure}

For the photon energy range of $10$--$15\unitev$, it is possible to
obtain a resonance process, and thus the ARPES intensity for the
A~$\rightarrow$~B~$\rightarrow$~D transition [see again
Fig.~\ref{Fig.2}(a)] is $10$ times larger than that for the
A~$\rightarrow$~C~$\rightarrow$~D transition.  In this case, the first
step of the A~$\rightarrow$~B~$\rightarrow$~D transition is the direct
optical transition, A~$\rightarrow$~B, from the carbon $\pi$ band to
the conduction bands around the $K$ point.  For this purpose, in
Fig.~\ref{dip}, we show the absolute value of the dipole vector,
$\mathbf{D}(\mathbf{k})=\langle
m\mathbf{k}|\nabla|i\mathbf{k}\rangle$, as a function of the
intermediate state energy for different conduction bands in graphite
[Figs.~\ref{dip}(a) and (b)] and graphene [Figs.~\ref{dip}(c) and
(d)].  For the initial states that satisfy Table II, we plot the
dipole vectors for $|i\rangle=B_2$ [Figs.~\ref{dip}(a) and (c)] and
for $|i\rangle=A_2$ [Figs.~\ref{dip}(b) and (d)].  The wave vector of
the initial state of the electron is considered at a point with a
distance of $2\pi \slash a\times 10^{-4}$ from the $K$ point along
the $\Gamma$--$K$ line.  The full circles, dots, and asterisks denote the
$x$-, $y$-, and $z$- components of the dipole vectors, i.e., $D_x$,
$D_y$, and $D_z$, respectively.  The symmetry of each intermediate
state is also labeled.

More detailed information about the dipole vectors plotted in
Figs.~\ref{dip}(a)-(d) can be obtained by comparing them with the
electronic band structures in Figs.~\ref{Fig.2}(a)-(b).  The two
lowest energy optical transitions around $E_m\approx 1\unitev$ shown
in Figs.~\ref{dip}(a) correspond to the $\pi\rightarrow \pi^*$
transitions of graphite.  Next, the optical transition around
$E_m\approx 11\unitev$ may originally correspond to the
$\pi\rightarrow B_2$ or the $\pi\rightarrow A_2$ transition, since
either choice is possible following Fig.~\ref{Fig.2}(a).  However, the
$\pi\rightarrow A_2$ optical transition can be excluded by the
selection rule in Table~\ref{tab:chartable2}.  At $E_m\approx
12\unitev$, the intermediate state can be the $A_1$, or to the $\sigma^*_{1}$
or $\sigma^*_{2}$ bands (see Fig.~2).  The nonzero value of the dipole
vector corresponds to $D_y$ for the $B_2\rightarrow A_1$ transition,
while the dipole vector becomes $\textbf{D}=0$ for $A_2\rightarrow
A_1$.  The $\sigma^*_{1}$ and $\sigma^*_{2}$ bands as the intermediate
states have $A_1$ and $B_1$ symmetries.  The dipole vector for the
$B_2\rightarrow B_1$ and $A_2\rightarrow B_1$ transitions are
$\textbf{D}=0$ and $D_y$, respectively.

In Figs.~\ref{dip}(c) and (d), we show similar properties with those
in Figs.~\ref{dip}(a) and (b), but now for the case of graphene.  The
direction of the dipole vector for the $\pi\rightarrow \pi^*$
transition at $E_m\approx 0\unitev$ along the $\mathit{\Gamma}$--$K$
and $K$--$M$ directions is $D_x$, which is consistent with the results
from Gr\"uneis \emph{et al.}~\cite{gruneis03}. The $\pi\rightarrow
B_2$ transition takes place at $E_m\approx11\unitev$ and the dipole
vector components for this transition are $D_z$ and $D_x$, as shown in
Figs.~\ref{dip}(c) and (d).  In the case of the third and fourth
lowest conduction bands in graphene, the band with $A_1$ symmetry
(orange dots in Fig.~\ref{Fig.2} around the $K$ point) and
$\sigma^*_1$ band with $A_1$ symmetry are both involved in the
electron-photon excitation.  The directions of the dipole vector for
the $\pi\rightarrow A_1$, and $\pi \rightarrow \sigma^*_{1}$
transitions along the $\Gamma$--$K$ and $K$--$M$ directions are
denoted by $D_y$ and $D_x$, respectively.

To discuss the magnitude of the dipole vectors for a given transition,
we project the wave function (plane wave) on the atomic wave
functions~\cite{silkin09,kogan14}.  The calculations show that the
$\pi~\rightarrow~B_2$ transition has the largest dipole vector among
the available transitions in Fig.~\ref{dip} since the $\pi$ and $B_2$
bands have the same $p_z$ orbital shape.  The
$\pi~\rightarrow~\sigma^*_1$ transition has the smallest dipole vector
because the $\sigma^*_1$ band near the $K$ point is formed by $p_x$
and $p_y$ orbitals which do not overlap with $p_z$.  On the other
hand, the $\pi \rightarrow \sigma^*_2$ and $\pi \rightarrow A_1$
transitions should also be taken into account because the $\sigma^*_2$
and $A_1$ bands are formed by the $s$, $p_x$, and $p_y$ orbitals.  The
dipole vectors for the $\pi \rightarrow \sigma^*_2$ and $\pi
\rightarrow A_1$ transitions are however weaker than that for the $\pi
\rightarrow B_2$ transition.

\begin{figure}
\includegraphics[width=80mm]{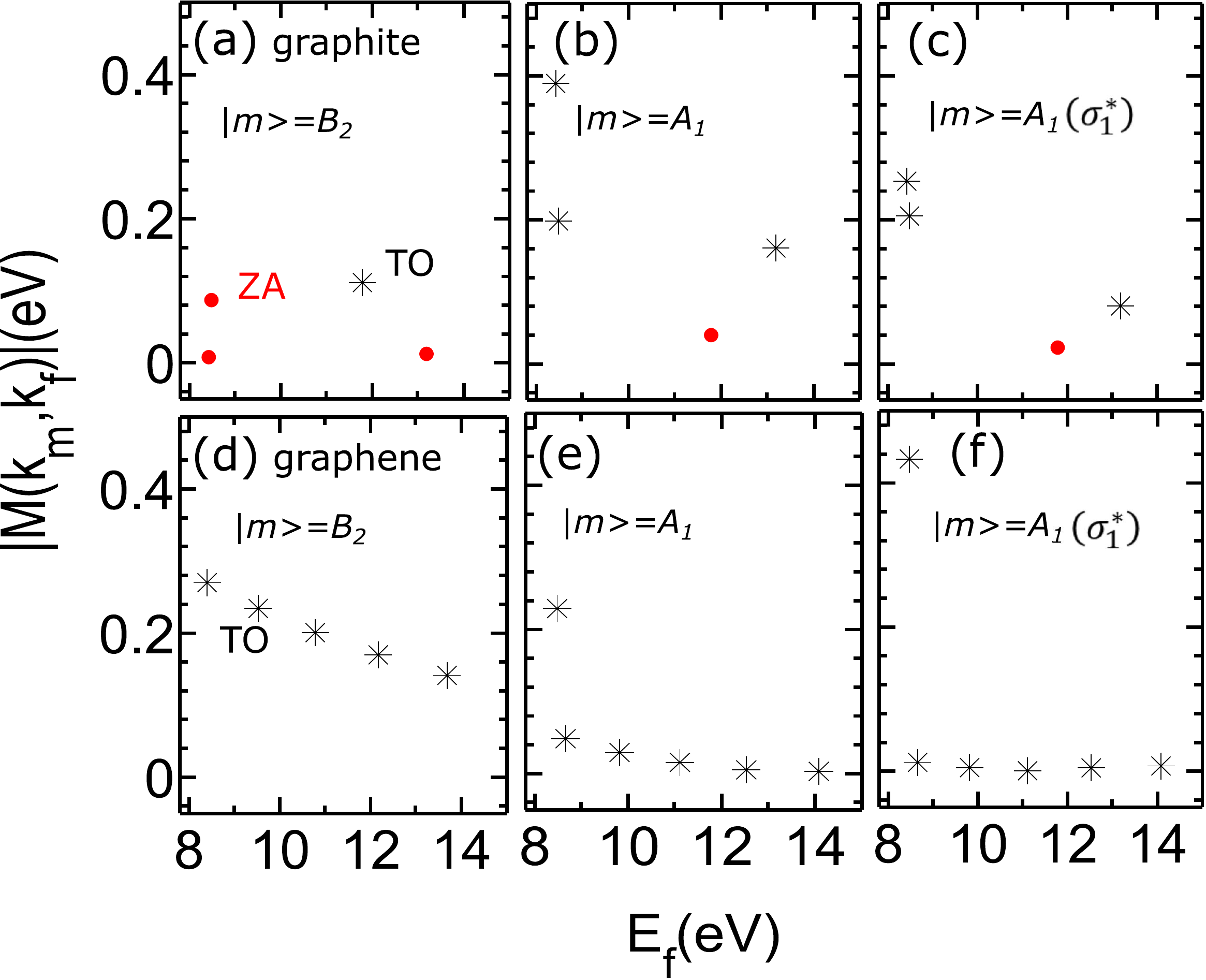}
\caption{\label{epcr} (Color online) The electron-phonon matrix
  elements for the scattering events from the intermediate states
  $|m\rangle$ with particular symmetries ($B_2$ and $A_1$) into some
  final states with different energies $E_f$.  Panels (a)-(c) are for
  graphite, while panels (d)-(f) are for graphene.  The dots and
  asterisks refer to the electron-phonon interaction for the ZA and TO
  phonon modes, respectively.  Note that $|m\rangle$ in panels (b)-(c)
  and (e)-(f) have the same symmetries but originate from different
  bands.  In particular, (c) and (f) are related with $|m\rangle$ of
  the $\sigma_1^*$ band.}
\end{figure}

The electron-phonon matrix element calculation reveals that, although
the TO and LA phonon modes have the same symmetry ($A_1$), the matrix
element for the LA phonon mode near the $K$ point is negligibly small.
The insignificant electron-phonon interaction for the LA phonon mode
near the $K$ point physically originates from the direction of atomic
displacements of the LA mode.  In Figs.~\ref{epcr}(a)-(f), we show the
calculated electron-phonon matrix elements as a function of the final
state energy, $E_f$, in graphite and graphene.  The dots and asterisks
correspond to the coupling of the photoexcited electron with the ZA
and TO phonon modes, respectively.  The difference between graphene
and graphite is physically related to the ZA phonon mode, which cannot
(can) be observed in the ARPES spectra for graphene (graphite),
because graphene does not have an interlayer electron-phonon
interaction~\cite{sato11a}.  Besides, the value of the electron-phonon
matrix element decreases with increasing $E_f$.

For the incident photon with $\hbar\omega\approx 11.1\unitev$,
photoexcited electrons in the $B_2$ band are scattered into the final
states near the $\Gamma$ point (see Fig.~2).  In the case of graphite,
the final states can be $\pi^*$, $\sigma^*_{1}$ or $\sigma^*_{2}$, as
intensity for the $\hbar\omega\approx 11.1\unitev$ arises from the
coupling between the photoexcited electron and the TO phonon.  From
these facts, we can conclude that the ARPES intensity around
$154\unitmev$ for $\hbar\omega\approx11.1\unitev$ is due to the
photoexcitation of an electron from the $\pi$ band to the $B_2$ band
which is then the scattered by the TO phonon mode into a state near
the $\Gamma$ point.  It should be noted that the discrepancy between
the experimental and theoretical binding energy might come from the
effect of the electron-electron correlation on the phonon
dispersion~\cite{lazzeri08b}, which is beyond the scope of this work.
For $\hbar\omega\approx 13\unitev$, the intermediate state can be
associated with the $A_1$, $\sigma^*_{1}$ and or $\sigma^*_{2}$ bands.
In this case, both the ZA and TO phonon modes can be coupled with the
photoexcited electron.  However, the electron-phonon interaction for
the ZA phonon mode is weaker than that for the TO phonon mode as
discussed above.  Thus, the ARPES intensity observed for
$\hbar\omega\approx12.5\unitev$ is assigned to the TO and ZA phonon
modes.

\subsection{Nonresonant indirect transition}

Now we consider the case when the incident photon energy is
$\hbar\omega\approx 6\unitev$.  The excitation process is the
nonresonant indirect transition and the final state is the $A_1$ band,
which is a nearly free-electron state.  Let us again discuss the
possibilities of the A~$\rightarrow$~B~$\rightarrow$~D and
A~$\rightarrow$~C~$\rightarrow$~D transitions.  If we assume that the
virtual state comes from the closest real states of the electrons, the
optical excitation in the second process
($\sigma_1,\sigma_2\rightarrow A_1$) has a negligible intensity
\cite{marinopoulos04}. Furthermore, the optical transition along the
high symmetry points on the $\Gamma$--$A$ line ( perpendicular to the
$\Gamma$--$K$--$M$--$\Gamma$ plane) for the second process also has a
negligible intensity.  Thus, the dominant mechanism should be the
$A\rightarrow B \rightarrow D$ transition.  As we mentioned before,
although we find that the A~$\rightarrow$~B~$\rightarrow$~D transition
would also be more preferable for $\hbar\omega\approx 6\unitev$, the
physical argument for why this transition is dominant for
$\hbar\omega\approx 6\unitev$ is different from that for
$\hbar\omega\approx 11.1\unitev$.

\begin{figure}
\includegraphics[width=65mm]{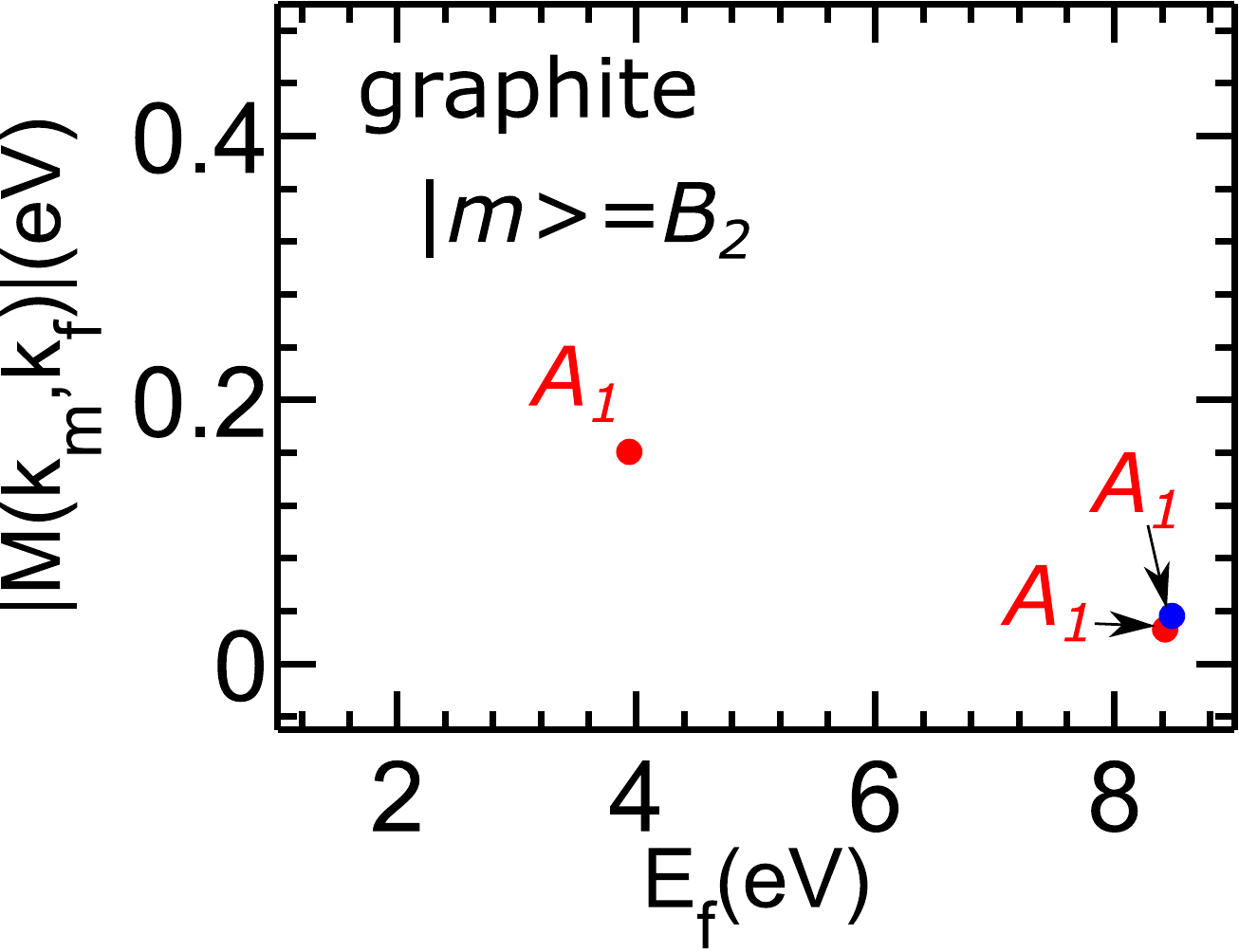}
\caption{\label{epcza} (Color online) Electron-phonon matrix elements
  for the ZA phonon mode of graphite for the transitions from an
  intermediate state $m\rangle$ having $B_2$ symmetry (the $\pi^*$
  band near the $K$ point) into some different final states with
  energies $E_f$.  }
\end{figure}

We can see that for the $A\rightarrow B \rightarrow D$ transition with
$\hbar\omega\approx 6\unitev$, the intermediate state is the $B_2$
band and the dominant dipole vector is $D_z$ (see Fig.~\ref{dip}).
Therefore, only the ZA phonon mode can be involved in this process
(see Table~\ref{tab:chartable2}).  The electron-phonon matrix element
as a function of the final state is plotted in Fig.~\ref{epcza}.  It
can be seen that there is a strong coupling between the $\pi^*$ band
and the $A_1$ band. The observation of the strong electron-phonon
coupling between the $\pi^*$ and $A_1$ bands was also reported with
scanning tunneling spectroscopy by Zhang \emph{et al.}  and Wehling
\emph{et al.}~\cite{zhang08g,wehling08}.  We conclude that the ZA
phonon mode corresponds to the ARPES signal at $E_b=67\unitmev$ if
photons with $\hbar\omega\approx 6\unitev$ and $p$-polarization are
incident on the graphite surface.
 
\begin{table}[t]
  \caption {Phonon ($|Ph\rangle$) assignment for different photon
    energies ($\hbar\omega$).  Columns for
    $|i\rangle$, $|m\rangle$ and $|f\rangle$ show the orbital shapes for 
    initial, intermediate, and final states, respectively, while
    $|O\rangle$ denotes the direction of the dipole vector.}
\label{tab:orb} 
\begin{tabular}{ c c c c c c c  c }
  \toprule
  $\hbar\omega(\unitev)$&{ } { }{ }$|i\rangle$ & { } { }{ }
  $|O\rangle$
  & { } { }{ } $|m\rangle$ & { } { }{ }$|Ph\rangle$
  &{   }{  }{  }$|f\rangle$ &   \\
  \hline
  $6$& $p_z$   & $D_z$  &  {}{    } $p_z$        & $ZA$     
  &{  } {  } $s$ \\ 
  $11$& $p_z$   &  $D_z$,$D_y$    &  {}{    } $p_z$        &  TO         &{  } {
  }$p_z$ \\ 

  $13$& $p_z$   &  $D_y$    &  $s,p_x,p_y$      & TO       &{  } {  }$s$     \\ 
  \toprule
\end{tabular}
\end{table}

When we look at the ARPES intensity for $\hbar\omega\approx 6\unitev$
and $\hbar\omega\approx 12.5\unitev$ in Fig.~5, there is a discrepancy
between the experimental data of the ARPES intensity and the
calculated results.  The experimental ARPES intensity is higher than
the calculated intensity for $\hbar\omega\approx 6\unitev$, while the
experimental ARPES intensity is much smaller than the calculated
intensity for $\hbar\omega\approx 12.5\unitev$.  The origin of these
discrepancies might be explained by the angle between the emission
direction of the ejected photoelectron and the
detector~\cite{liu10aa}.  The direction of the detector is considered
to be normal to the surface in the experiment~\cite{liu10aa,tanaka13}
(see Fig.~\ref{symm}).  In Table~\ref{tab:orb}, we show the shapes of
the orbitals for the initial state $|i\rangle$, the intermediate state
$|m\rangle$, the final state $|f\rangle$, the dipole vector direction,
$|O\rangle$, and the phonon polarization, $|Ph\rangle$, for photon
energies $\hbar\omega=6\unitev$, $11\unitev$ and $13\unitev$.  Every
initial state is in the $\pi$ electron band, formed by the $p_z$
orbital. 

For the $\hbar\omega\approx6\unitev$ transition, the $|m\rangle$ state
also has the $p_z$ orbital character.  The dipole vector becomes $D_z$
and the out-of-plane phonon mode ZA also couples to the photoexcited
electron. In this case, the final state, $|f\rangle$, has an $s$
orbital shape.  Therefore, the ejected electron from this excitation
process can be observed in the direction normal to surface more
strongly.  For the $\hbar\omega\approx11\unitev$ excitation, the
$|m\rangle$ state also has a $p_z$ shape and the dipole vector also
becomes $D_z$.  However, in this case, the electrons couple to the
in-plane phonon mode TO and $|f\rangle$ has a $p_z$ shape.  As a
result, the ejected electrons from this process also can be
well-observed in the direction normal to the surface.  But we should
note that the intensity of the observed electrons can decrease due to
the coupling between the electron and the phonon mode.  For the
$\hbar\omega\approx12.5\unitev$, the intermediate state has
$s,p_x,p_y$ orbital shapes and the dipole vector is $D_y$ and also the
electron is coupled with the in-plane TO phonon mode, and in this case
the final state has an $s$ orbital shape.  The ejected electrons from
this process thus have a large dipole vector component parallel to the
surface so that the possibility of the observation of the electrons
from this process when the detector is normal to the surface will
dramatically decrease.

\begin{figure}
\includegraphics[width=80mm]{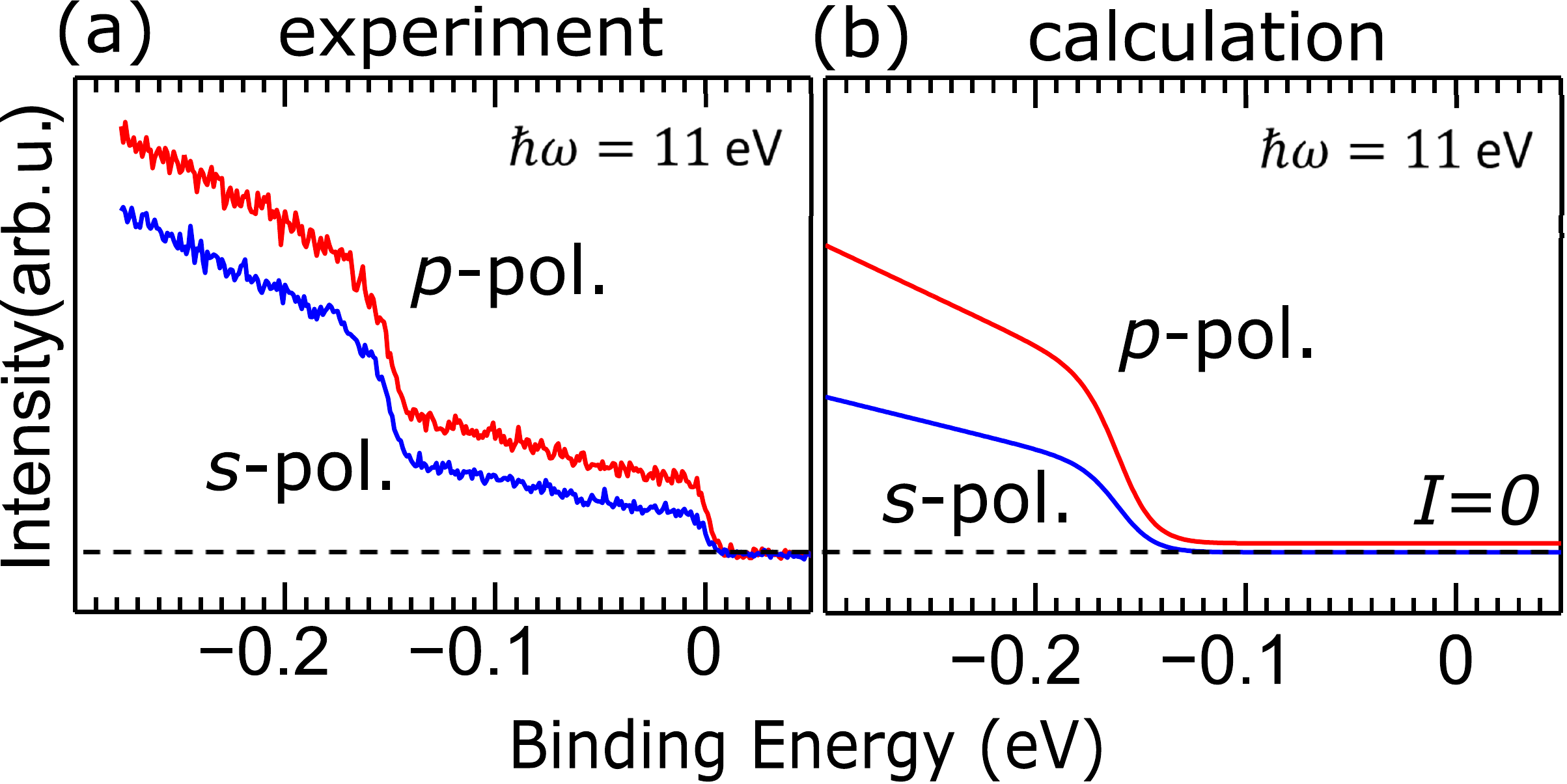}
\caption{\label{epc-ps} (Color online) (a) Experimental measurement
  and (b) theoretical calculation of the ARPES intensity as a function
  of the binding energy for a single crystal of graphite.  The energy
  of the incident photon is $\hbar\omega=11\unitev$.  The top and
  bottom curves correspond to the ARPES intensity for $s$-polarized
  and $p$-polarized light, respectively.}
\end{figure}

\subsection{Effects of $s$- and $p$-polarizations}

Finally, we discuss the polarization dependence of the incident light
for a single-crystal of graphite.  In Figs.~\ref{epc-ps}(a) and (b),
we plot the experimental and calculation data of graphite for
$s$-polarized and for $p$-polarized light in the case of
$\hbar\omega=11\unitev$.  It can be seen that the ARPES intensity for
$p$-polarized light is stronger than $s$-polarized light for both
experimental measurement and theoretical calculation.  The physical
reason for this behavior is that the $z$-component of the vector
potential ($A_z$) is stronger than the $x-$component ($A_x$) for
$\hbar\omega=11\unitev$ and $\psi=40^{\circ}$, although the dipole
vector components $D_z$ and $D_x$ have the same magnitude~(see
Fig.~\ref{dip}).  Note that there is a small jump at $E_b=0$ observed
experimentally, originating from the phonon absorption in the indirect
transition \cite{liu10aa}, that we did not consider in the
calculations.

\section{Summary} 

The indirect transition for the ARPES spectra in graphene and graphite
have been investigated for different incident photon energies and
light polarizations.  We have compared the theoretical calculation of
the indirect transition for the ARPES intensity of graphene and
graphite with experimental measurements for HOPG and graphite.  Our
symmetry analysis shows that the ZA, TO, and LA phonon modes, which
have even symmetry with respect to the mirror plane,
$\sigma_v^\prime(yz)$, can be involved in the indirect interband
transition.  Although the LA phonon mode has even symmetry with
respect to the mirror plane, its phonon energy cannot be observed
because it has a negligible electron-phonon interaction near the $K$
point in the Brillouin zone.  Thus, the ARPES spectra with binding
energy $E_b=154\unitmev$ is assigned to the TO phonon modes of
graphene and graphite when $p$-polarized photons with
$\hbar\omega\approx11\unitev$ are used.  The relevant mechanism for
the observation of the TO phonon mode is a resonant indirect
transition.  Meanwhile, for the incident photons with
$\hbar\omega\approx6\unitev$, the ZA mode becomes dominant, being
observable through a nonresonant indirect transition occurring in
graphite for $p$-polarized light.  Therefore, the phonon branches
which can be observed by the ARPES measurement have been here assigned
based on the detailed symmetry analysis and calculations, which were
not available in the previous experiment~\cite{tanaka13}.
Furthermore, the ARPES intensity of graphite for $p$-polarized light
is stronger than for $s$-polarized light when the incident photon
energy is $\hbar\omega\approx11\unitev$ because the vector potential
of the $p$-polarized light is expected to be stronger than that of
$s$-polarized light.

By understanding the indirect transitions in the ARPES spectra of
graphite and graphene, we expect that more detailed phonon dispersion
relations might be observed in our future experiments.  Besides, we
believe that the validity of our methods should not be limited to
graphene-based materials.  We propose that the electron-phonon
coupling for a large class of two-dimensional materials should also be
observable by ARPES with indirect transitions.

\section*{Acknowledgments}
The ARPES experiments were performed under the UVSOR joint studies
program of the Institute for Molecular Science (UVSOR, IMS), and under
the aproval of the Hiroshima Synchrotron Radiation Center (HiSOR,
Proposal No.13-B-9). S.T. expresses sincere thanks to Dr. S. Ideta and
Prof. K. Tanaka of UVSOR and to Mr. S. Arita and Prof. K. Shimada of
HiSOR for their kind assistance. S.T. thanks all the members of the
UVSOR and HiSOR facilities for their valuable help during the
experiments.  P.A. is supported by the MEXT scholarship.
A.R.T.N. acknowledges the financial support from the Leading Graduate
School Program in Tohoku University.  R.S. acknowledges MEXT Grants
Nos. 25107005 and 25286005. M.S.D. acknowledges NSF grant
No. DMR-1507806.

\appendix

\section{Electron-phonon interaction}
\label{sec:epc}
Let us define the equilibrium position of an atom $\sigma=A,B$ in the
$n$th unit cell by $\mathbf{R}^n_\sigma$
\begin{equation}
\mathbf{R}^n_\sigma=\mathbf{R}_n+\mathbf{d}_\sigma
\label{eq:position}
\end{equation}
where $\mathbf{R_n}$ and $d_\sigma$ are, respectively, positions of
the unit cell and the relative position of the $\sigma$~th atom in the
unit cell.
 
The changes of the potential energy due to the lattice displacement are given by
\begin{align}
  H_{\rm ep} & = \sum_{n,\sigma}
  [V_n(\mathbf{r}-\mathbf{R}^n_\sigma+S^{\alpha}_{n,\sigma}(t))-
  V_n(\mathbf{r}-\mathbf{R}^n_\sigma)] \notag \\
  & =\sum _{n,\sigma} S^{\alpha}_{n,\sigma}(t) \cdot
  \nabla_{\mathbf{R}^n_\sigma} V_n(\mathbf{r}-\mathbf{R}^n_\sigma),
\end{align}
in which $S^{\alpha}_{n,\sigma}(t)$ denotes the displacement vector of the atom 
and $\alpha=1,\ldots,6$ denotes the phonon mode, where
\begin{equation}
  S^{\alpha}_{n,\sigma}(t)=A^{\alpha}_{\rho}(\mathbf{q})e^{\alpha}_\sigma(\mathbf{q})
  e^{i\mathbf{q}.\mathbf{R}^n}e^{\pm i\omega^{\alpha}(\mathbf{q})t}
\end{equation}
where $A^{\alpha}_{\rho}$ is the 
amplitude of the atomic vibration. The $\pm$ sign and
$\rho$ indices refer to whether a phonon is emitted ($"-"$ and $\rho =
E$) or absorbed ($"+"$ and $\rho = A$). Here, $e^{\alpha}(\mathbf{q})$ is
the unit vector of the lattice displacement vector, and
$\omega(\mathbf{q})$ is the angular frequency of the phonon with a wave
vector $\mathbf{q}$.  The amplitude of the vibration,
$A^{\alpha}_{\rho}$, is given by
\begin{equation}
  A^{\alpha}_{\rho}(\mathbf{q})=\sqrt{\frac{2\hbar N^{\alpha}_{\rho}
      (\mathbf{q})}{m\omega^{\alpha}(\mathbf{q})N}}
\end{equation}
where $N^{\alpha}_{\rho}$ denotes the number of the phonons for
the $\alpha$-th phonon mode and $N$ is the number of atoms in the sample
that contribute to the phonon, $m=1.9927 \times 10^{-26}$ kg is the
mass of a carbon atom. $N^{\alpha}_{A}$ and $N^{\alpha}_{E}$ are given
by the Bose-Einstein distribution function as follows:
\begin{align}
  N^{\alpha}_A(q) &=\frac{1}{\exp(\frac{\hbar\omega^{\alpha}
      (\mathbf{q})}{k_BT})-1},\notag\\
  N^{\alpha}_E(q)&=N^{\alpha}_A(q)+1.
\end{align}

Here, we adopt the rigid-ion approximation where the potential $V$
rigidly follows the motion of the
ions~\cite{grimvall80,glembocki82}. Thus, the electron-phonon
interaction Hamiltonian can be expressed by
\begin{equation}
\begin{split}
  H_{\rm ep} = -\sum_{n=0}^{N-1}\sum_{\sigma=A,B}\sum_{\alpha=1}^6
  &A_{\rho}^\alpha(\mathbf{q})
  S^\alpha_{n,\sigma}(t) \cdot
  \nabla_{\mathbf{r}}
  V_n(\mathbf{r}-\mathbf{R}^{n,\alpha}_\sigma),
\label{eq:hep}
\end{split}
\end{equation}
where we adopted the fact that $\nabla_r V_n=\nabla_R V_n$.  Using 
perturbation theory, the nonzero matrix elements for this potential
are given by
\begin{equation}
  M^{v,v'}_{\rm ep}(\mathbf{k}_f,\mathbf{k}_i)=
  \langle \mathbf{k_f}|H_\text{ep}|\mathbf{k_i} \rangle,
\label{eq:matrix}
\end{equation}
where $v$ and $v'$ label the initial and final states.

To calculate the electron-phonon matrix elements,
we expand the wave function of the initial states and final
states in terms of plane waves,
\begin{equation}
\begin{split}
  |\mathbf{k}_i^v\rangle & =\frac{1}{\sqrt{V}}
  \sum_{\mathbf{G}}C^{i,v}_\mathbf{G}
  (\mathbf{k}_i)\exp\Big({i\big(\mathbf{k}_i+\mathbf{G}\big)
    \cdot \mathbf{r}\Big)}, \\
  |\mathbf{k}_f^{v'}\rangle &
  =\frac{1}{\sqrt{V}}\sum_{\mathbf{G'}}C^{f,v'}_\mathbf{G'}
  (\mathbf{k}_f) \exp\Big({i\big(\mathbf{k}_f+\mathbf{G'}\big)\cdot
    \mathbf{r}}\Big),
\label{eq:planewave}
\end{split}
\end{equation}
where $V$ is the volume of the sample, $\mathbf{G}$ represents the
reciprocal lattice vector of graphene and
$C_{\mathbf{G}}^{i,v}(C_{\mathbf{G}}^{f,v'})$ are the plane-wave
coefficients.  Thus, the electron-phonon matrix elements are given by
\begin{equation}
\begin{split}
  M^{v,v'}_{\rm ep}(\mathbf{k}_f,\mathbf{k}_i)& = \frac{1}{V}
  \sum_{n=0}^{N-1}\sum_{\alpha=1}^6
  \sum_{\sigma=A,B}\sum_{\mathbf{G,G'}}C_{\mathbf{G'}}^{*f,v'}(\mathbf{k}_f)
  C_{\mathbf{G}}^{i,v}(\mathbf{k}_i)\\
  & \times A^{\alpha}_{\rho}(\mathbf{q})e^{i\mathbf{q}\cdot
    \mathbf{R}^n}e^{\alpha}_\sigma(\mathbf{q})\cdot
  \mathbf{m}_D(\mathbf{k_f},\mathbf{k_i}),
\label{eq:matrix2}
\end{split}
\end{equation}
where $\mathbf{m}_D$ is expressed by  
\begin{equation}
\begin{split}
\mathbf{m}_D(\mathbf{k}_f,\mathbf{k}_i)& = \int
e^{i(\mathbf{k}_f-\mathbf{k}_i+\mathbf{G'}-
\mathbf{G})\cdot
\mathbf{r}}\nabla_{\mathbf{r}}V(\mathbf{r}-\mathbf{R}_{\sigma}^n)
d\mathbf{r}.  
\label{eq:atomicm}
\end{split}
\end{equation}
Then, by changing variables
$\mathbf{r}'=\mathbf{r}-\mathbf{R}_{\sigma}^n$,
and $d\mathbf{r}'=d\mathbf{r}$,  
$\mathbf{m}'_D(\mathbf{k}_f,\mathbf{k}_i)$ is expressed by
\begin{equation}
\begin{split}
\mathbf{m}'_D(\mathbf{k}_f,\mathbf{k}_i)& = \int
e^{i(\mathbf{k}_f-\mathbf{k}_i+\mathbf{G'}-\mathbf{G})
\cdot \mathbf{r}'}\nabla_{\mathbf{r}'}V(\mathbf{r}')d\mathbf{r}'.  
\label{eq:atomicm2}
\end{split}
\end{equation}
To obtain an explicit expression for the electron-phonon matrix
element, we multiply the following unity relation into
Eq.~\eqref{eq:matrix2}:
\begin{equation}
\begin{split}
1 = e^{i(\mathbf{k}_f-\mathbf{k}_i+\mathbf{G'}-\mathbf{G})
\cdot \mathbf{R}_{\sigma}^n} e^{-i(\mathbf{k}_f+\mathbf{G'})\cdot \mathbf{R}_{\sigma}^n}
e^{i(\mathbf{k}_i+\mathbf{G})\cdot \mathbf{R}_{\sigma}^n}.
\label{eq:one1}
\end{split}
\end{equation}
the electron-phonon matrix elements are given by
\begin{align}
  M^{v,v'}_\text{ep}&(\mathbf{k}_f,\mathbf{k}_i)\notag\\
  =& \frac{1}{V} \sum_{n=0}^{N-1}\sum_{\alpha=1}^6
  \sum_{\sigma=A,B}\sum_{\mathbf{G,G'}}C_{\mathbf{G'}}^{*f,v'}
  (\mathbf{k}_f)C_{\mathbf{G}}^{i,v}(\mathbf{k}_i)\notag\\
  &\times A^{\alpha}_{\rho}(\mathbf{q})e^{-i(\mathbf{k}_f-\mathbf{k}_i
    +\mathbf{G'}-\mathbf{G}) \cdot \mathbf{R}_{\sigma}^n}
  e^{i\mathbf{q}\cdot \mathbf{R}^n} \notag\\
  &\times e^{\alpha}_\sigma(\mathbf{q})\cdot
  \mathbf{m}'_D(\mathbf{k}_f,\mathbf{k}_i).
\label{eq:matrix3}
\end{align}

\begin{table}[t!]
  \caption {Coefficients $v_p$ and
    $\tau_p$ for the ion potential $V(\mathbf{r})$ in
    Eq.~\eqref{eq:potential}~\cite{jiangp04,jiang05}.  The units of
    $v_p$ are Hartree $\times$ a.u.,
    and $\tau_p$ is in atomic units. ($1$ Hartree is $27.211$ eV and 
    $1$ a.u. is $0.529177\unitangstrom$) }
\label{tab:coeff} 
\begin{tabular}{ r | r r r r r r  r }
  \toprule
  p{  }& 1 & {  } 2 & { } 3 & { } 4
    \\
  \hline
  $v_p$ {  }& {  }-2.13 {  }  &  -1.00 {  }    &  -2.00  {  }       & -0.74        \\ 

  $\tau_p$ {  }& 0.25 {  }   &  0.04   {  }  &  1.00   {  }    & 2.80           \\ 
     \toprule
\end{tabular}
\end{table}

Using the fact that $\sum_{n=0}^{N-1}e^{-i(\mathbf{k}_f-\mathbf{k}_i
  -\mathbf{q}+\mathbf{G'}-\mathbf{G}) \cdot \mathbf{R}^n}=
\delta_{\mathbf{k}_f,\mathbf{k}_i+\mathbf{q}}$ in
Eq.~(\ref{eq:matrix3}) and using Eq.~(\ref{eq:position}), we get the
following electron-phonon matrix element:
\begin{align}
  M^{v,v'}_{\rm ep} &(\mathbf{k}_f,\mathbf{k}_i) \notag\\
  =&\frac{1}{V} \sum_{\alpha=1}^6
  \sum_{\sigma=A,B}\sum_{\mathbf{G,G'}}C_{\mathbf{G'}}^{*f,v'}(\mathbf{k}_f)
  C_{\mathbf{G}}^{i,v}(\mathbf{k}_i) \notag\\
  &\times A^{\alpha}_{\rho}(\mathbf{q})e^{-i(\mathbf{k}_f-\mathbf{k}_i
    +\mathbf{G'}-\mathbf{G}) \cdot \mathbf{d}_{\sigma}}\notag\\
  &\times\delta_{\mathbf{k}_f,\mathbf{k}_i+\mathbf{q}}
  e^{\alpha}_\sigma(\mathbf{q})\cdot
  \mathbf{m}'_D(\mathbf{k}_f,\mathbf{k}_i).
\label{eq:matrix4}
\end{align}

In order to obtain Eq.~(\ref{eq:atomicm2}), we expand an ion potential,
$V(\mathbf{r})$, of a free carbon atom, obtained by the ab-initio method
\cite{jiangp04,jiang05}, into a sum of Gaussian basis functions as
follows:
\begin{equation}
\begin{split}
V(\mathbf{r})=-\frac{1}{r}\sum_{p=1}^4v_p\exp{(\frac{-\mathbf{r}^2}{2\tau^2_p})}. 
\label{eq:potential}
\end{split}
\end{equation}      
The fitting parameters for the potential in Eq.~\eqref{eq:potential}
are listed in Table~\ref{tab:coeff}.

Finally, putting Eq.~\eqref{eq:potential} into the
Eq.~\eqref{eq:atomicm2}, we get
$\mathbf{m}'_D(\mathbf{k_f},\mathbf{k_i})$ as follows
\begin{align}
  \mathbf{m}'_D(\mathbf{k_f},\mathbf{k_i})=&
  -i2\pi\sqrt{2\pi}\frac{\mathbf{Q}}{|\mathbf{Q}|}
  \sum_{p=1}^4v_p\tau_p\text{Erfi}
  \left[\frac{|\mathbf{Q}|\tau_p}{\sqrt{2}}\right] \notag\\
  &\times\exp{\left[-\left(\frac{|\mathbf{Q}|\tau_p}{\sqrt{2}}\right)^2\right]}
\label{eq:atomicm3}
\end{align}
where $\mathbf{Q}=\mathbf{q}+\mathbf{G'}-\mathbf{G}$ and $\text{Erfi}(z)$ is the
imaginary error function and it is defined as
\begin{align}
\text{Erfi}(z)=-i\text{Erf}(iz),
\label{eq:erf}
\end{align}
where $\text{Erf}(z)$ is defined by
\begin{align}
\text{Erf}(z)= \frac{2}{\sqrt{\pi}}\int_0^z e^{-t^{2}}dt.
\label{eq:erf1}
\end{align}

\section{Basic ARPES mechanism and
the lattice symmetry}
\label{sec:lsym}

\begin{figure}
  \centering\includegraphics[width=8cm]{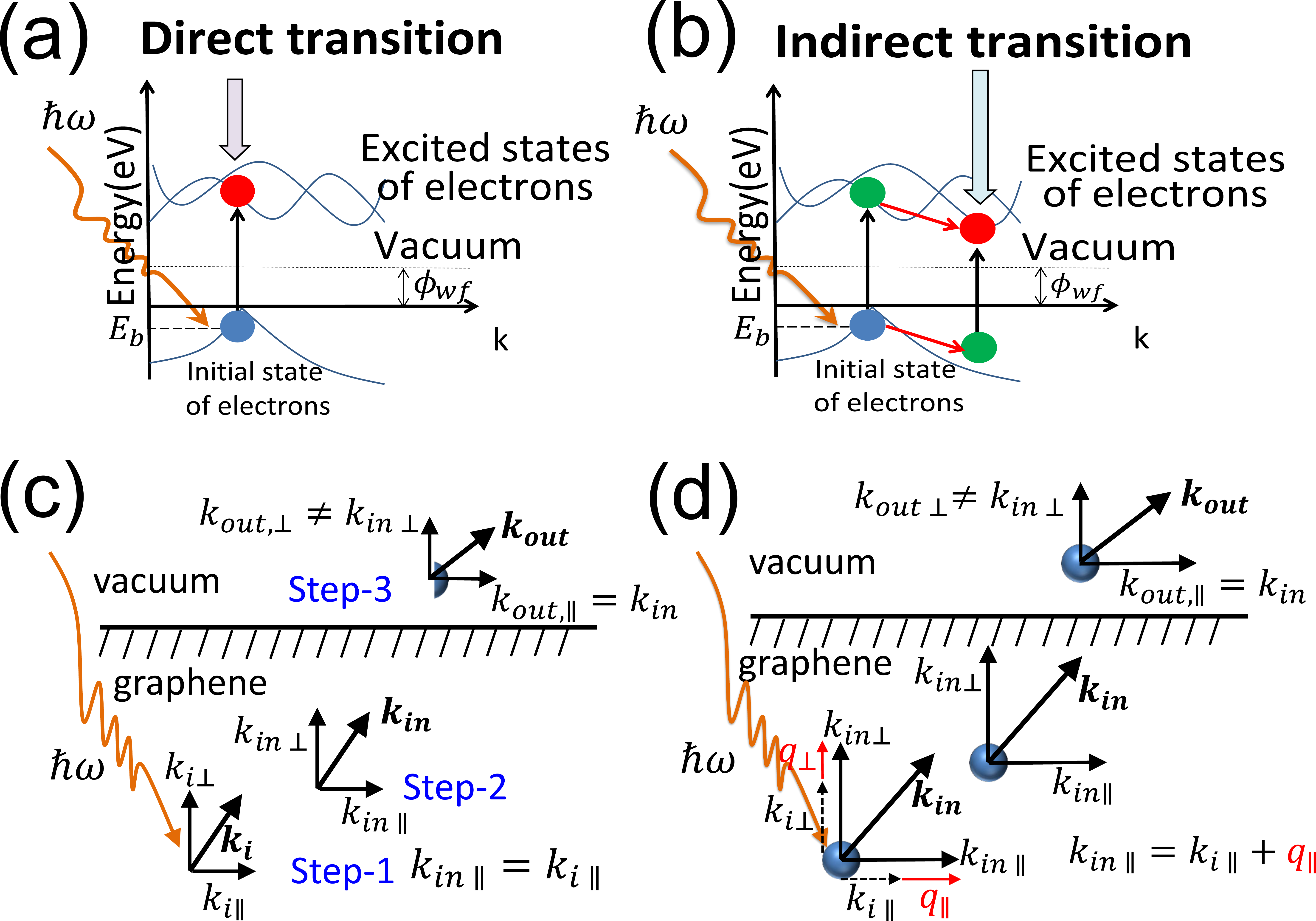}
  \caption{\label{fig:arpes}(Color online) Schematic representations of
    (a,c) direct and (b,d) indirect transitions.  In the direct
    (indirect) transition, the momenta of the electron before and
    after the transition are the same (different). Panel (c) shows a
    photoemission process in the three-step model for the direct
    transition~\cite{suga14}. Panel (d) shows a photoemission process
    in the three-step model for the indirect transition. The
    difference between the direct and indirect transitions is on the
    first step, where the momentum of the phonon $q$ is added to the
    momentum of the electron in the indirect transition.}
\end{figure}

ARPES is one of spectroscopy methods to observe the electronic
dispersion relations of the occupied bands of
solids~\cite{damascelli04}.  In the case of the direct transition,
this process can be divided into three steps, the so-called three-step
model: (1) photoexcitation of an electron inside the solid [see
Fig.~\ref{fig:arpes}(a)], (2) travel of the photoelectron to the solid
surface by an incident momentum [see Fig.~\ref{fig:arpes}(a)], and (3)
emission of the photoelectron into the vacuum [see
Fig.~\ref{fig:arpes}(c)].

The photoemission intensity as a function of the binding energy and
momentum of the electron shows the electron dispersion relations of
solids.  The binding energies of the electron inside the sample and
outside the sample, respectively, are determined by the energy
conservation rules:
\begin{equation}
 E_{\rm kin,in} = \hbar\omega-|E_b|
\end{equation}
and
\begin{equation}
 E_{\rm kin,out} = \hbar\omega-\phi-|E_b|
\end{equation}
where $E_{\rm kin,in}$ and $E_{\rm kin,out}$ are the kinetic energy of
electron inside and outside of the sample, respectively, $\hbar \omega$ is
the photon energy, $\phi$ is the work function of the solid and
$|E_b|$ is the binding energy.

The momentum of the electron parallel to the surface of the solid is
conserved during this process because the force is applied to the
photoelectron only in the perpendicular direction to the
surface~\cite{suga14}, as shown in Fig.~\ref{fig:arpes} (c).  Thus,
the parallel momentum of the electron inside the sample is related to
the parallel momentum of the electron outside of the sample as
follows:
\begin{equation}
  k_{{\rm out},\parallel}=k_{{\rm in},\parallel}
\end{equation}

In the case of the indirect transition [see Fig.~\ref{fig:arpes}(b)],
although the three-step model is still appropriate, the first step of
energy and momentum conservation has an additional term [see
Fig.~\ref{fig:arpes}(d)], which come from the momentum of the phonon.
The energy conservation outside of the sample is written as follows
\begin{equation}
 E_{\rm kin,in} = \hbar\omega-\phi-|E_b + E_{b,q}|
\end{equation}
where $E_{b,q}$ expresses the binding energy of the electron after
scattering and
\begin{equation}
  k_{{\rm out},\parallel}=k_{{\rm in},\parallel}=k_{i,\parallel}+q_{\parallel}
\end{equation}
where $q_{\parallel}$ is the additional momentum that the electron
absorbs after the electron-phonon scattering.

In the case of graphene-based materials and similar materials such as
silicene and germanene, the highest binding energy of the electron is
limited at the Fermi level ($E_b=0$).  Therefore, the energy
conservation of the electron is reduced to
\begin{equation}
 E_{\rm kin} = \hbar\omega-\phi-|E_{b,q}|.
\end{equation}   
Then, resolving the energy of the electron after scattering becomes
possible.  Furthermore, in the indirect transition, since the
observation of the electron is near the $\Gamma$ point ($k_{{\rm
    out},\parallel}=0$), and the initial state of the electron is
limited to be near the $k_{{\rm in},\parallel}=K$ point, the momentum
conservation of the electron in this process is satisfied when
$q_{\parallel}=K$.

\begin{figure}
  \centering\includegraphics[width=8cm]{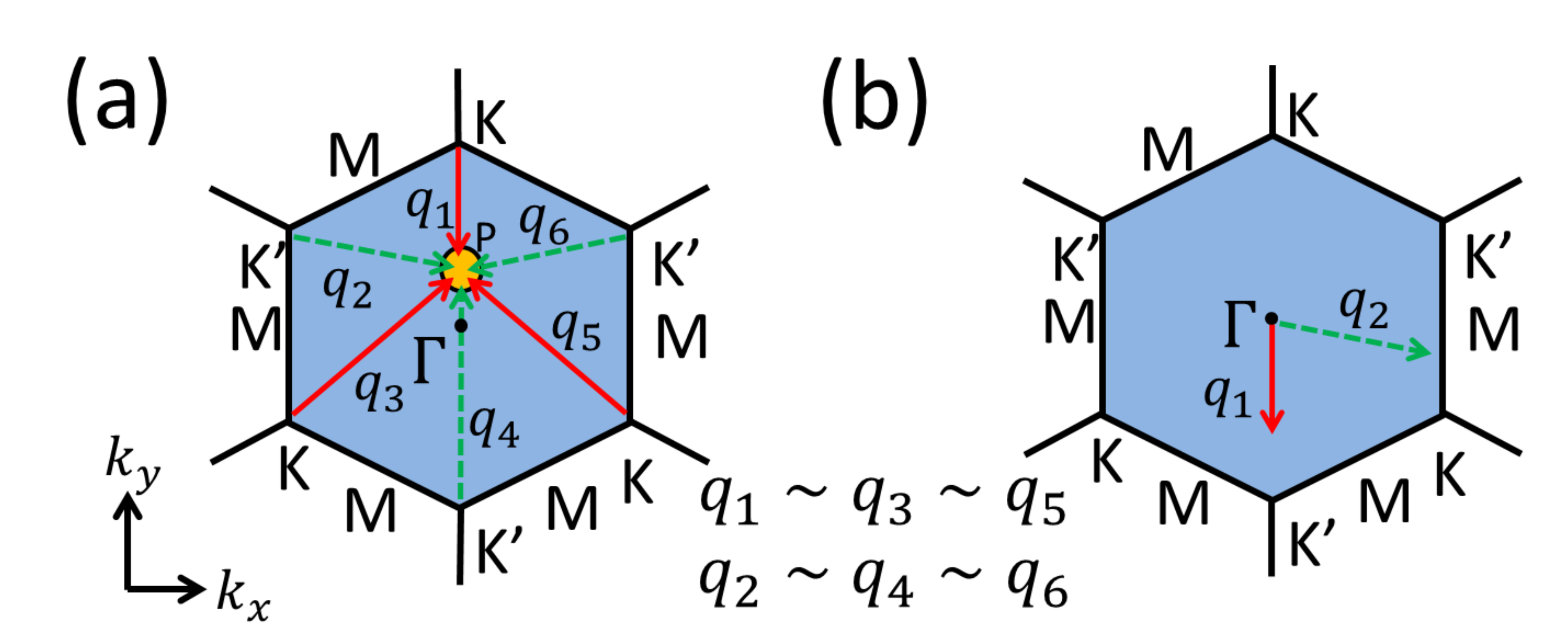}
  \caption{\label{ss1}(Color online) (a) Electrons are scattered from
    around the $K$ or $K'$ point into a $\mathbf{k}$ point, labeled P, shown
    in (a) as a yellow dot, along $\Gamma$--$K$ by phonons.  The phonon
    wave vectors which scatter electrons from $K$ and $K'$ into the
    observation point, P, are shown by red and green dot arrows,
    respectively.  The phonon wave vector
    $\mathbf{q}_1,\mathbf{q}_3,\mathbf{q}_5$ and
    $\mathbf{q}_2,\mathbf{q}_4,\mathbf{q}_6$ indicate scattering from
    $K$ and $K'$ respectively.  (b) Only two phonon momenta are
    inequivalent, whereas the phonon momenta
    $\mathbf{q}_3,\mathbf{q}_5$ and $\mathbf{q}_4,\mathbf{q}_6$ are
    folded back into the first Brillouin Zone, $\mathbf{q}_1$ and
    $\mathbf{q}_2$, respectively, due to the lattice
    symmetry~\cite{liu10aa}.}
\end{figure}

Now, we consider that electrons are scattered by phonons from around
the $K$ or $K'$ points into a certain $\mathbf{k}$ as shown by a
yellow circle at the $\Gamma$--$K$ line in Fig.~\ref{ss1}~(a).  The
phonon wave vectors which scatter electrons from the $K$ and $K'$
points into $\mathbf{k}$, are shown by the red solid and green dot
arrows, respectively.  The phonon wave vectors
$\mathbf{q}_1,\mathbf{q}_3,\mathbf{q}_5$ and
$\mathbf{q}_2,\mathbf{q}_4,\mathbf{q}_6$ indicate scattering from the
$K$ and $K'$ points, respectively.  However, only two phonon momenta
are inequivalent whereas the phonon momenta
$\mathbf{q}_3,\mathbf{q}_5$ and $\mathbf{q}_4,\mathbf{q}_6$ are folded
into the first Brillouin zone $\mathbf{q}_1$ and $\mathbf{q}_2$,
respectively, due to the lattice symmetry~\cite{liu10aa}, see
Fig.~\ref{ss1}~(b).  As a result, when the ARPES intensity along the
$\mathit{\Gamma}$--$K$ direction is investigated, the phonons
$\mathit{\Gamma}$--$K'$ and $K$--$M$--$K'$ can also be observed.
Similarly, when the ARPES intensity along the $\mathit{\Gamma}$--$K'$
direction is investigated, the phonon along the $\mathit{\Gamma}$--$K$
and $K'$--$M$--$K$ directions can be observed.  Thus, we can
distinguish whether the electrons scatter from the $K$ or $K'$ points
by observation of the $\mathit{\Gamma}$--$K'$ or $K$--$M$--$K'$ phonon
dispersions, respectively.


\end{document}